\documentclass[lettersize,journal,comsoc]{IEEEtran}
\usepackage{geometry}
\IEEEoverridecommandlockouts
\usepackage{cite}
\usepackage{graphicx}
\usepackage{subfigure}
\usepackage{color}
\usepackage{bm}
\usepackage[cmex10]{amsmath}
\usepackage{array}

\usepackage{algorithm}
\usepackage{algpseudocode}  
\usepackage{multirow}
\newcommand{\tabincell}[2]{\begin{tabular}{@{}#1@{}}#2\end{tabular}} 

\usepackage{paralist}
\usepackage{setspace}
\usepackage{stfloats}
\usepackage{array,color}
\usepackage{enumerate}
\usepackage{threeparttable}

\usepackage{amsthm}
\usepackage{amssymb}
\newtheorem{theorem}{Theorem}
\newtheorem{lemma}{Lemma}

\begin{document}




\title{Task-Oriented Image Semantic Communication Based on Rate-Distortion Theory}
\author{\IEEEauthorblockN{Fangfang Liu, \textit{Member}, \textit{IEEE}, Wanjie Tong, Yang Yang, \textit{Member, IEEE}, Zhengfen Sun, and Caili Guo, \textit{Senior Member, IEEE}}}



\maketitle

\renewcommand{\thefootnote}{\fnsymbol{footnote}}

\footnotetext{Fangfang Liu, Wanjie Tong, Yang Yang, and Zhengfen Sun are with the Beijing Key Laboratory of Network System Architecture and Convergence, School of Information and Communication Engineering, Beijing University of Posts and Telecommunications, Beijing 100876, China (e-mail: fliu@bupt.edu.cn; twj@bupt.edu.cn; young0607@bupt.edu.cn; sunzhengfen06@bupt.edu.cn).

Caili Guo is with the Beijing Laboratory of Advanced Information Net works, School of Information and Communication Engineering, Beijing University of Posts and Telecommunications, Beijing 100876, China (e-mail: guocaili@bupt.edu.cn).}

\begin{abstract}
Task-oriented image semantic communication is a new communication paradigm, which aims to transmit semantics for artificial intelligent (AI) tasks while ignoring the reconstruction quality of the images. However, in some applications, such as autonomous driving, both image reconstruction quality and the performance of the followed AI tasks must be simultaneously considered. To tackle this challenge, this paper proposes a task-oriented semantic communication scheme with semantic reconstruction (TOSC-SR). Its main goal is to simultaneously minimize pixel-level and task-relevant semantic-level distortion during communications under a certain rate, which formulates a new rate-distortion optimization problem. To successfully measure the loss at the semantic level, a new form of semantic distortion measured by the mutual information between the semantic-reconstructed images and the task labels is proposed. Then, we derive an analytical solution for the formulated problem, where the self-consistent equations of the problem are obtained to determine the optimal mapping of the source and the semantic-reconstructed images. To implement TOSC-SR, we further obtain an extended form of rate-distortion form based on the variational approximation of mutual information, which is applicable to multiple AI tasks. Experimental results show that the proposed approach outperforms the traditional JPEG, JPEG2000, BPG, VVC-based image communication systems and deep learning based benchmarks in terms of image reconstruction quality, AI task performance, and multi-task generalization ability.

\end{abstract}


\IEEEpeerreviewmaketitle
\section{Introduction}
\IEEEPARstart{C}{ommunication} technologies, such as 5G, have achieved great success in the past years. However, as the capacity gradually approaches the Shannon capacity limits, the existing communications technologies that aim at transmitting bits as massive and fast as possible is doomed to face a bottleneck due to the limited resource. Under this background,  semantic communications, a higher level of communication envisioned  by Shannon and Weaver, has attracted increasing attention. The goal of semantic communication is semantic information exchange rather than the transmission of symbols \cite{shannon1948mathematical, strinati20216g}. With the tighter and deeper integration of communications and artificial intelligence (AI), semantic communication is unfolding as a revolutionary paradigm of the next generation wireless networks \cite{strinati20216g, xie2021deep}. In this novel paradigm, not only the wireless transceivers are spanning from normal devices to intelligent entities with edge intelligence capabilities, but also the applications are extended from connecting people to connecting intelligent unmanned systems with cloud intelligence capabilities. Besides human, semantic communications can also empower the intelligent entities to accomplish various AI tasks with much higher spectral efficiency, especially for visual tasks like pedestrian monitoring, defect detecting, and security surveillance \cite{SPIC}. To this end, the semantic information of the images must be efficiently compressed and transmitted.

In particular, existing studies adopt the entropy of the compressed representation and the quality of the received image to evaluate the rate and the distortion performance of an image communication system, respectively\cite{Balle}. To balance the rate and distortion performance, a rate-distortion optimization problem is typically formulated. In particular, the distortion measurement is generally defined by the distance between source and its compressed representation according to the rate-distortion theory \cite{cover1999elements}. 
According to different communication goals, distortion measurement can be divided into three categories: the pixel-level distortion, the feature-level distortion, and the semantic-level distortion.

Conventional image communication typically aims to minimize the pixel-level distortion. Here the pixel-level distortion is generally quantified by the mean-squared-error (MSE) between the original input images and the reconstructed images in pixels, and performance metrics such as peak signal-to-noise ratio (PSNR) are typically used as the performance measure. As the core of image communication, existing image compressions rely on the handcrafted time-frequency transforms, such as JPEG \cite{JPEG}, JPEG 2000 \cite{JPEG2000}, BPG \cite{BPG} based on HEVC\cite{HEVC}, and VVC \cite{VVC}, or the deep learning based methods, including auto-encoder (AE) and generative adversarial networks (GANs) \cite{Balle, Theis, Toderici, ScaleHyperprior, CAECheng}. These methods are mainly committed to achieving a better trade-off between the rate and the average reconstruction distortion of the images. Though minimizing the pixel-level distortion can be applied to various AI tasks, the performance of the AI tasks and the reconstruction quality of the images may not be always linearly correlated. Therefore, more works have been proposed to ahiceve better AI task performance.

One specific type is image communication aiming to minimize the feature-level distortion \cite{DIC}. In these works, the feature distortion is measured by the distance between the features extracted from the input source images and that extracted from the reconstructed images. The main purpose of this type of image communications is to reserve the global features of source in the reconstructed images. Even though the downstream tasks can be well accomplished, these works do not directly take the performance of the AI tasks into consideration.

To directly optimize downstream AI task performance, image communication systems aiming to minimize the semantic-level distortion have been proposed. In these works, only the semantic information related to the considered specific AI task is extracted and transmitted \cite{IB1999, xie2021deep, SCAITasks, SPIC, DeepSIC}. Different from those minimizing pixel-level or feature-level distortions, the core design problem of this type becomes to capture the semantics of the considered task's goal \cite{SPIC, DeepSIC}. In particular, information bottleneck (IB) distortion can be used to measure the semantic distortion of an AI task by measuring the distance of the semantic information for the target task. As indicated in \cite{SPIC, DeepSIC, qin2021survey}, different tasks lead to different semantic features even for the same source images. Moreover, the smaller IB distortion implies the more relevant semantic features for the considered specific task, which, however, can degrade the generalization ability of the system to perform other AI tasks. 

From above analysis, we can conclude that a task oriented semantic communication scheme that achieves both superior reconstruction quality and performance for general AI tasks still deserves investigation. To tackle this challenge, this work develops a semantic communication scheme that minimizes the pixel-level distortion and semantic-level distortion while reserving the generality for various AI tasks. This design problem is formulated as an extended rate-distortion problem. To our best knowledge, this is the first work on task-oriented semantic reconstruction from the rate-distortion optimization perspective. The main contributions of this paper are summarized as follows:

\begin{itemize}
\item We define a new extended rate-distortion form to guide the semantic representation, transmission, and reconstruction for multiple AI tasks. In particular, we first establish a new distortion measurement, which consists of both pixel-level distortion and semantic-level distortion. Based on this extended distortion measurement, we formulate the image semantic transmission problem as an extended rate-distortion optimization problem.

\item We solve the formulated problem by introducing a Lagrange multiplier, $\beta$, for the semantic-level distortion. Then, We derive a closed-form analytical solution for the extended rate-distortion optimization problem, where the self-consistent equations are obtained to determine the optimal mapping of source and the semantic-reconstructed images by taking target task into account. 

\item  Considering practical applications, we further propose an approximation form of the extended rate-distortion that is applciable to DNN based task-oriented semantic communication systems. In partiuclar, to obtain a differentiable optimized goal for DNN's learning process, we derive a relaxed version of the extended rate-distortion based on variational approximation. We further analyze the semantic reconstruction from the perspective of information theory. To estimate the mutual information between high dimensional latent features, we further apply a feasible mutual information estimator contrastive log-ratio upper bound (CLUB) method \cite{CLUB}.




\end{itemize}


The rest of this paper is organized as follows. Related works are reviewed in Section II. The system model is presented and a corresponding problem is formulated in Section III. Section IV describes the proposed TOSC-SR scheme. The experimental results along with discussions are presented in Section V. Finally, conclusions are drawn in Section VI.

\section{Related Work}

In this section, we first review the definition and measurement of semantic information over the past few decades, and then introduce the various semantic communication frameworks. The related image compression techniques, and the theoretical guidance, including rate distortion and information bottleneck of the communication, are also reviewed.


\subsection{Definition and Measurement of Semantic Information}
The definition and measurement of semantic information has always been a key issue. In classic information theory, anything that could reduce uncertainty is regraded as information and therefore entropy is used as the measure of information according to its statistical characteristics\cite{shannon1948mathematical}. Analogously, earlier researchers have also defined semantic information for textual processing based on probability. Slightly different from the classic information theory, Carnap et al.\cite{Carnap} developed the semantic information theory based on logical probabilities ranging over the contents. Barwise et al. \cite{Situation} further proposed the principle of scene logic to define semantic information. Floridi et al.\cite{Strong} introduced the theory of strongly semantic information. Since fuzzy concepts are very common in text communication and different people have different understandings, Zadeh used fuzzy sets and fuzzy events to describe the fuzziness of semantic information, which is measured by membership function \cite{Zadeh_1, Zadeh_2}. However, due to different subjective semantic understandings of the same message by different people, it is 
challenging to obtain the logical probability and membership functions accurately in practice.

Recently, most researchers in semantic communication adopt deep neural networks (DNNs) to define, extract, and measure semantic information\cite{IB1999, xie2021deep, 6GSemantic, Speech_Recognition}. In these works, semantics is generally considered as the features extracted from the source that are related to a specific task. This definition of semantic information can be traced back to the IB theory \cite{IB1999, IB2015}, where the mutual information was used to measure semantic information about the task in the source or in the extracted features. It is very tricky to calculate mutual information between high-dimensional variables. Fortunately, many scalable and flexible mutual information estimation methods based on DNNs have been proposed in recent studies\cite{MINE, CLUB}. By taking advantage of DNNs, it is feasible to obtain the amount of semantic information by means of mutual information between the raw data and the task labels or between the reconstructed signals and the task labels, which provides the basic foundation for our work. 

\begin{figure*}[t]
    \centering
    \includegraphics[scale=0.75]{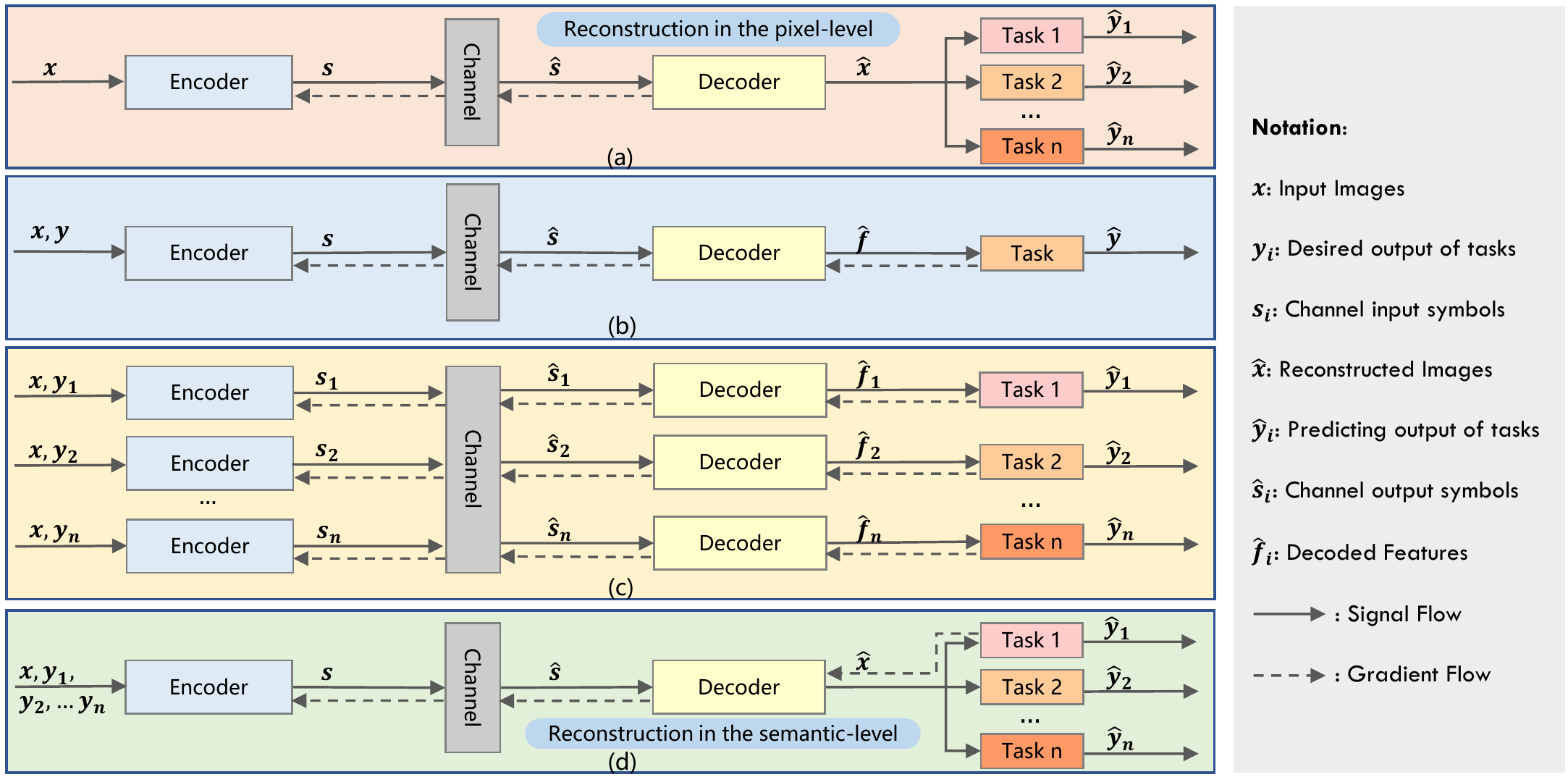}
    \caption{Comparisons of different image communication frameworks with AI tasks: (a) general communication for image reconstruction in the pixel or feature domain \cite{JSCC, JSCC-f, DIC, 6GSemantic}, (b) semantic communication for one specific task \cite{JointTR, SCAITasks, JSCCRetrieval, ImageRetrieval, multi_modal_1}, (c) semantic communication for multiple tasks with respective codecs \cite{DeepSIC, SPIC, multi_modal_2}, and (d) our proposed semantic communication for multiple tasks with one generalized codec.}
    \label{fig_comparison_frameworks}
\end{figure*}

\subsection{Frameworks of Semantic Communication Systems}
With the help of DNNs, a plethora art of semantic communication schemes has been proposed based on Transformer\cite{xie2021deep, multi_modal_2} and CNN\cite{SPIC, Semantic_speech, 6GSemantic, JointTR, SCAITasks, JSCCRetrieval, ImageRetrieval, JSCC, JSCC-f, DIC, DeepSIC, multi_modal_1}. According to different communication goals, the image semantic communication frameworks can be classified in four types, as shown in Fig. \ref{fig_comparison_frameworks}: 1) general image communication for image reconstruction in the pixel or feature level, 2) image semantic communication for one specific task, 3) image semantic communication for multiple tasks with respective codecs, and 4) image semantic communication for multiple tasks with one generalized codec by reconstructing the images with semantics, and the proposed work belongs to this type. The encoder and decoder shown in Fig. \ref{fig_comparison_frameworks} can adopt the joint source-channel codec or separated source-channel codec \cite{JSCC}.

As shown in Fig. \ref{fig_comparison_frameworks}(a), the purpose of this type of image communication is to reconstruct the original image using the extracted features in the pixel-level, so that the recovered signal and the original signal are as consistent as possible in appearance. Bourtsoulatze et al. \cite{JSCC} designed a JSCC image communication system based on CNN, where PSNR is used to measure the accuracy of image recovery at the receiver. By incorporating the channel output feedback into the transmission system, Kurka et al. \cite{JSCC-f} improved the reconstruction quality at the receiver. In order to make reconstruted images similar to original images in both pixel-level and feature-level, the communication system in \cite{strinati20216g} jointly optimizes the appearance loss and multi-scale discriminator loss. Though the reconstructed images are more similar to the original ones in both appearance and perception, the whole reconstruction training process does not involve the gradient feedback of tasks, so the semantic information relevant to task is still not considered.

As shown in Fig. \ref{fig_comparison_frameworks}(b), the purpose of general image communication is to reconstruct image in the pixel-level while that of semantic communication is to use the extracted semantic information to directly complete the specific task at the receiver, without the need to reconstruct the original image.  The joint transmission-recognition in \cite{JointTR} performs the feature extraction and recognition at the IoT device and server, respectively. Along this line, the transmitted data is further compressed in \cite{SCAITasks} by introducing the semantic relationship between feature maps and semantic concept.  A task-based compression scheme was proposed in \cite{JSCCRetrieval, ImageRetrieval} for the image retrieval task in the wireless edge scenario. The task-oriented multi-user semantic communication system for multimodal data transmission, named MU-DeepSC, can execute the visual question answering(VQA) task. In summary, these non-reconstructed based semantic communication schemes often accompany with specific tasks, which may lose the generalization ability for the other tasks.

As shown in Fig. \ref{fig_comparison_frameworks}(c), this type of image semantic communication system considers multiple tasks at the receiver. By incorporating the semantic information into the codec during image compression, deep semantic compression (DeepSIC) was introduced in \cite{DeepSIC}, where two semantic compression networks was designed to perform the semantic analysis after the feature extractor in the transmitter. In line with this idea, the modified auto-encoder networks in \cite{SPIC} have better multi-task (classification and reconstruction) performance. The schemes in \cite{DeepSIC, SPIC} address multi-task type semantic communication systems, in which the multiple tasks will compete with each other for better performance, resulting a suboptimal solution. In order to unify the structure of transmitters for different tasks, a Transformer based unique framework was proposed in \cite{multi_modal_2}.

Motivated by the existing semantic communication frameworks, we propose a task-oriented semantic communication system with semantic reconstruction, as shown in Fig. \ref{fig_comparison_frameworks}(d). Here the input images are reconstructed in the semantic-level to not only maintain the task performances with the gradient feedback but also enhance the generalization ability among different tasks. In this way, the transmission cost can be effectively reduced by using the semantic-reconstructed images to accomplish the various tasks at the receiver.

\subsection{Image Compression and Semantic Representation}
Images are playing an increasingly important role in current communication systems. Different from the natural language, which is highly semantic and information-dense \cite{he2021masked}, images are generally natural signals with quite objective descriptions. The rich semantic information is contained in images with heavy spatial redundancy, which leads image compression a critic problem.

Traditional lossy image compression methods, such as JPEG, JPEG2000, BPG based on HEVC, and VVC, rely on hand-crafted module design individually. Each module was designed with multiple modes and optimized by the rate-distortion theory to determine the best mode. Recently, a great number of learning-based image compression models utilizing auto-encoder architecture have been proposed, which have achieved great success with promising results\cite{Balle, Theis, Toderici, ScaleHyperprior, CAECheng}. Similar to the traditional methods, these models were still optimized to minimize the reconstruction distortion, without considering the contents or semantics of images. The content-weighted strategy in \cite{ImportantMap} allocates more bits to the important parts of an image, which was determined by a simple fully connected network. Considering the region of interest (ROI) of an image, the end-to-end optimized ROI image compression scheme in \cite{ROI} achieves better visual quality and compression performance than the traditional compression methods in ROI. The existing compression schemes take no task performance into consideration, so the features of image compression are in essence general representations for image reconstruction without semantic consideration.

To extract more semantic information related to the AI tasks during image compression, semantic presentations have been studied extensively. By exploiting DNNs, intelligent entities can perform various AI tasks, such as image classification, image generation, semantic segmentation, object detection, image retrieval, and so on \cite{DeepVision}. As different tasks require different semantic representations, how to obtain effective semantic representations with limited wireless resource becomes a very worthwhile problem.

\subsection{Rate Distortion and Information Bottleneck}


Existing works on lossy image compression\cite{Balle, Theis, ScaleHyperprior} and communication\cite{JSCC, JSCC-f} are based on the rate-distortion theory, where the rate is defined as the mutual information between source and its representation, and the distortion is directly defined as the difference between the reconstructed images and the original images in the pixel or feature domain. 

As an extension of the rate-distortion theory, the IB theory was first proposed by Thshby et al. \cite{IB1999} in 1999. A theoretic framework finds concise representations related to the task variable for the source variable. The IB distortion is defined and extended as the mutual information between representation and task labels. The IB theory has been applied to various supervised and unsupervised tasks \cite{App1, App2, App3}.  Recently, it has been used to explain the optimization process of DNNs \cite{IB2015}. For instance, the visual experiments in \cite{IB2017} showed that the optimization process is indeed corresponding to the optimization objective of IB. Inspired by this, a variational approximation in \cite{VIB} was adopted to the IB, then a supervised learning model constructed by DNN using the relaxed IB trade-off was optimized, which can obtained better performance than other forms of regularization.

Though some preliminary semantic communication systems have been proposed by researchers, to our best knowledge, there is limited theoretical research investigates semantic communications based on the rate-distortion theory. In this work, we further extend the rate-distortion theory by exploiting its generalization ability among different tasks. In particular, a new semantic distortion is used as the trade-off between the semantic reconstructed distortion and the IB distortion relevant to AI tasks.

\section{Problem Formulation}
\begin{figure*}[t]
\centering
\includegraphics[scale=0.60]{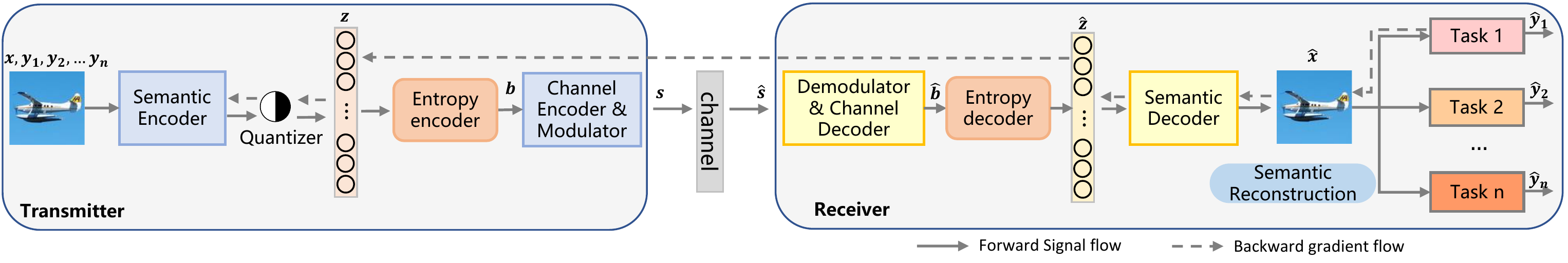}
\caption{System model of the TOSC-SR scheme.}
\label{fig_Semantic_image_transmission_framework}
\end{figure*}

We consider a point-to-point image semantic communications between an IoT device and the edge server. As shown in the Fig. \ref{fig_Semantic_image_transmission_framework}, the transmitter is an IoT device collecting images while the receiver is the edge server to receive the transmitted image for further processing. The target of this work is to achieve appropriate trade-off between pixel-level reconstruction distortion and task-relevant distortion between original images and reconstructed images so as to improve the reconstruction quality and AI task performance. In this section, we introduce the details of the considered image semantic communications model and formulate it as an extended rate-distortion optimization problem.

\subsection{System model}
As shown in the Fig. \ref{fig_Semantic_image_transmission_framework}, the proposed image semantic communication system is composed of a transmitter, a channel, and a receiver. The receiver may need to accomplish a multiple AI tasks according to the specific requirements of the system. In particular, the input image, $\bm{x} \in \mathbb{R}^{C\times H\times W}$, is first transformed to $\bm{e}$ by semantic encoder for effective compression and transmission, obtaining the representation
\begin{equation}
    \bm{e} = S_{\theta}(\bm{x}),
\end{equation}
where $S_{\bm{\theta}}(\cdot)$ indicates the semantic encoder with parameters $\bm{\theta}$. To realize available entropy encoding, representation $\bm{e}$ must be quantized, which can be expressed as
\begin{equation}
    \bm{z} = Q(\bm{e}),
\end{equation}
where $Q(\cdot)$ is a quantizer. Then we convert these discrete representations to bit sequence by entropy encoder, obtaining
\begin{equation}
    \bm{b} = E(\bm{z}),
\end{equation}
where $E(\cdot)$ indicates the entropy encoder. To alleviate the affect of channel noise, channel coding and modulation are adopted following the source encoding, obtaining $k$ channel input symbols
\begin{equation}
    \bm{s} = M(C(\bm{b})),
\end{equation}
where $C(\cdot)$ indicates the channel encoder, and $M(\cdot)$ indicates the modulator. 

A single Rayleigh fading channel model is considered. Given the channel input symbols $\bm{s}\in \mathbb{R}^{k}$, the corresponding channel output symbols, $\hat{\bm{s}}$, can be given by
\begin{equation}
    \hat{\bm{s}} = h\bm{s}+\bm{n},
\end{equation}
where $h\in \mathbb{C}$, $\bm{n}\in \mathbb{C}^k$ are independent identically distributed (i.i.d) samples from the Gaussian distribution $\mathcal{N}(0,\sigma^{2}_{h} I_1)$ and $\mathcal{N}(0,\sigma^{2}_{\bm{n}} I_k)$ with variances $\sigma^{2}_{h}$ and $\sigma^{2}_{\bm{n}}$, respectively.


At the receiver, the noised symbols are firstly processed by the demodulator and the channel decoder, obtaining the recovered bit sequence
\begin{equation}
    \hat{\bm{b}} = C^{-1}(M^{-1}(\hat{\bm{s}})),
\end{equation}
where $M^{-1}(\cdot)$ indicates the demodulator, and $C^{-1}(\cdot)$ indicates the channel decoder. Subsequently, the semantic representations are recovered by the entropy decoder, obtaining
\begin{equation}
    \hat{\bm{z}} = E^{-1}(\hat{\bm{b}}),
\end{equation}
where $E^{-1}(\cdot)$ indicates the entropy decoder. Then $\hat{\bm{z}}$ is transformed back to pixels, obtaining the reconstruction image
\begin{equation}
    \hat{\bm{x}} = S^{-1}_{\bm{\phi}}(\hat{\bm{z}}),
\end{equation}
where $S^{-1}_{\bm{\phi}}(\cdot)$ indicates the CNN based semantic decoder with respect to parameters $\bm{\phi}$. We further input $\hat{\bm{x}}$ into the main AI task, producing the predicting output
\begin{equation}
    \hat{\bm{y}} = F_{\chi}(\hat{\bm{x}}),
\end{equation}
where $F_{\bm{\chi}}(\cdot)$ indicates the AI task network with respect to parameters $\bm{\chi}$. Benefiting from the cascade structure of image reconstruction network and AI task network, both the reconstruction distortion's loss and AI task's loss can produce gradient to instruct the whole network's training.

\subsection{Problem Formulation}
Denote $\bm{X}$ and $\hat{\bm{X}}$ as input image dataset and reconstructed image dataset, respectively, which means $\bm{x} \in \bm{X}$ and $\hat{\bm{x}} \in \hat{\bm{X}}$. We adopt the conventional MSE reconstruction distortion,
\begin{equation}\label{eq_MSE_loss_function}
    D_{R}(\bm{X},\hat{\bm{X}}) = \mathbb{E}_{p(\bm{x},\hat{\bm{x}})}||\bm{x}-\hat{\bm{x}}||^2,
\end{equation}
as the pixel-level distortion function.

The conventional rate-distortion optimization' goal is to minimize the pixel-level reconstruction distortion $D_R$ between $\bm{X}$ and $\hat{\bm{X}}$ at a given rate limit, which is represented as
\begin{equation}\label{eq_rate_distortion_optimizaiton}
    \underset{
        \substack{
        p(\hat{\bm{x}}|\bm{x}):
        I(\bm{X},\hat{\bm{X}}) \leq I_c
        }
    }
    {\min} D_{R}(\bm{X},\hat{\bm{X}}),
\end{equation}
where $I_c$ denotes information constraint and
\begin{equation}\label{eq_rate}
    R = I(\bm{X};\hat{\bm{X}}) = \sum_{\bm{x}} \sum_{\hat{\bm{x}}} {p(\bm{x},\hat{\bm{x}}) \log \left[ \frac{p(\bm{x},\hat{\bm{x}})}{p(\bm{x})p(\hat{\bm{x}})}  \right]}.
\end{equation}

In semantic communications, the goal becomes to minimize the trade-off between reconstruction distortion and AI task distortion in certain rate so as to improve the reconstruction quality and AI task performance. To this end, we propose a new form of distortion function, which trades off pixel-level reconstruction distortion $D_R$ and task-relevant distortion $D_{T}$ applied in the rate-distortion theory and information bottleneck principle, respectively. Based on this new distortion function, we formulate the image semantic communications as an extended rate-distortion optimization problem.

By introducing an additional task-relevant label set $\bm{Y}$, we can obtain the task-relevant distortion $d_{T}(\bm{x},\hat{\bm{x}})$ between input image $\bm{x}$ and reconstructed image $\hat{\bm{x}}$, which can be represented by KL divergence $D_{KL}[p(\bm{y}|\bm{x})|p(\bm{y}|\hat{\bm{x}})]$ \cite{IB2015}
\begin{equation}\label{eq_d_IB}
  d_{T}(\bm{x},\hat{\bm{x}}) = D_{KL}[p(\bm{y}|\bm{x})|p(\bm{y}|\hat{\bm{x}})] = \sum_{\bm{y}\in \bm{Y}} p(\bm{y}|\bm{x})\log \left(\frac{p(\bm{y}|\bm{x})}{p(\bm{y}|\hat{\bm{x}})} \right).
\end{equation}
Then the expectation of task-relevant distortion between $\bm{X}$ and $\hat{\bm{X}}$ can be represented by
\begin{small}
    \begin{equation}\label{eq_DIB_1}
    \begin{aligned}
         D_{T}(\bm{X},\hat{\bm{X}}) = \sum_{\bm{x}\in \bm{X}}\sum_{\hat{\bm{x}}\in \hat{\bm{X}}}\sum_{\bm{y}\in \bm{Y}}p\left(\bm{x},\hat{\bm{x}}\right)p\left(\bm{y}\middle|\bm{x}\right)\ \log{\left(\frac{p\left(\bm{y}\middle|\bm{x}\right)}{p\left(\bm{y}\middle|\hat{\bm{x}}\right)}\right)}.
    \end{aligned}
    \end{equation}
\end{small}

%

Though (\ref{eq_DIB_1}) provides a precise definition for task-relevant distortion, it is still challenging to calculate due to the lack of probability distributions of source data. The following lemma, proved in Appendix A, expresses that task-relevant distortion into another format, which is easier to understand and calculate.
\begin{lemma}\label{lemma_1}
The task-relevant distortion $D_{T}$ can also be represented as
\begin{equation}\label{eq_DIB_2}
    D_{T}(\bm{X},\hat{\bm{X}}) = I(\bm{X};\bm{Y}) - I(\hat{\bm{X}};\bm{Y}),
\end{equation}
which means the reduction amount of information about task label $\bm{Y}$ contained in reconstructed image $\hat{\bm{X}}$ compared to original image $\bm{X}$. The two definitions of task-relevant distortion are equivalent.
\end{lemma}

Taking both reconstruction distortion $D_R$ and task-relevant distortion $D_{T}$ into considerations, we can define the following distortion measurement,
\begin{equation}\label{eq_DS}
    D(\bm{X},\hat{\bm{X}})= D_{R}(\bm{X},\hat{\bm{X}}) + \beta D_{T}(\bm{X},\hat{\bm{X}}),
\end{equation}
where $\beta$ controls the trade-off between reconstruction distortion and task-relevant distortion.

Similar to conventional rate-distortion optimization, to reduce the communication overhead, we apply a constraint on the rate. Then we have the following optimization problem,
\begin{equation}\label{eq_optimization_2}
    \underset{
        \substack{
        p(\hat{\bm{x}}|\bm{x}):
        I(\bm{X},\hat{\bm{X}}) \leq I_c
        }
    }
    {\min} D_{R}(\bm{X},\hat{\bm{X}}) + \beta D_{T}(\bm{X},\hat{\bm{X}}).
\end{equation}


This optimization problem can be solved by the Lagrange multiplier method and the only variable is $p(\hat{\bm{x}}|\bm{x})$. The complete process for solving this problem can be expressed in Theorem 1 shown in Appendix B. 

\begin{theorem}\label{theorem_2}
The optimal mapping from source image set $\bm{X}$ to semantic-reconstructed image set $\hat{\bm{X}}$ should satisfy the following equation,
\begin{equation} \label{self_consistent_equation_03}
    \begin{aligned}
        p(\hat{\bm{x}}|\bm{x})=\frac{p(\hat{\bm{x}})e^{-\lambda ^{-1} d_{S}(\bm{x},\hat{\bm{x}})}}{\mu(\bm{x})},
    \end{aligned}
\end{equation}

\begin{equation} \label{self_consistent_equation_02}
    p(\hat{\bm{x}}) = \sum_{\bm{x}}{p(\bm{x})p(\hat{\bm{x}}|\bm{x})},
\end{equation}

\begin{equation} \label{self_consistent_equation_01}
    \begin{aligned}
        p(\bm{y}|\hat{\bm{x}}) &= \sum_{\bm{x}}{p(\bm{y}|\bm{x})p(\bm{x}|\hat{\bm{x}})},
    \end{aligned}
\end{equation}
where $\mu(\bm{x})$ is given by
\begin{equation}
    \mu(\bm{x})=\sum_{\hat{\bm{x}}}p(\hat{\bm{x}})e^{- \lambda^{-1} d_{S}(\bm{x},\hat{\bm{x}})}.
\end{equation}

\end{theorem}


To find the optimal distributions $p(\hat{\bm{x}}|\bm{x})$, $p(\hat{\bm{x}})$, and $p(\bm{y}|\hat{\bm{x}})$, it is natural to think of using alternate iteration Blahut–Arimoto algorithm\cite{cover1999elements}, according to the traditional rate distortion function. First, we pick a $\lambda$ and a $\beta$, and initialize distribution $p(\hat{\bm{x}})$ and $p(\bm{y}|\hat{\bm{x}})$, then minimize the semantic distortion to find the distribution $p(\hat{\bm{x}}|\bm{x})$. The self-consistent equations (\ref{self_consistent_equation_03}), (\ref{self_consistent_equation_02}), and (\ref{self_consistent_equation_01}) are satisfied simultaneously at the minima of the functional $\mathcal{L} (p(\hat{\bm{x}}|\bm{x}))$. The minimization is done independently over the convex sets of the normalized distributions, $p(\hat{\bm{x}})$ and $p(\bm{y}|\hat{\bm{x}})$ and $p(\hat{\bm{x}}|\bm{x})$. Namely
\begin{equation}
    \underset{p(\bm{y}|\hat{\bm{x}})}{\min} \underset{p(\hat{\bm{x}})}{\min} \underset{p(\hat{\bm{x}}|\bm{x})}{\min} \mathcal{L}(p(\hat{\bm{x}}|\bm{x});p(\hat{\bm{x}});p(\bm{y}|\hat{\bm{x}})).
\end{equation}
This minimization is performed by the converging alternating iterations. For the k-th iteration, we have the self-consistent equations as
\begin{equation} \label{eq_solution}
    \left\{
        \begin{aligned}
             p_k(\hat{\bm{x}}|\bm{x}) &= \frac{p_k(\hat{\bm{x}})e^{-\lambda^{-1} d_{S}(\bm{x},\hat{\bm{x}})}}{\sum_{\hat{\bm{x}}}p_k(\hat{\bm{x}})e^{-\lambda^{-1} d_{S}(\bm{x},\hat{\bm{x}})}},\\
        p_{k+1}(\hat{\bm{x}}) &= \sum_{\bm{x}}{p(\bm{x})p_k(\hat{\bm{x}}|\bm{x})},\\
        p_{k+1}(\bm{y}|\hat{\bm{x}}) &= \sum_{\bm{y}}p(\bm{y}|\bm{x})\frac{p_k(\hat{\bm{x}}|\bm{x})p(\bm{x})}{p_k(\hat{\bm{x}})}.
        \end{aligned}
    \right.
\end{equation}

It should be noted that $d_S(\bm{x}|\hat{\bm{x}})$ will directly affect the mapping $p(\hat{\bm{x}}|\bm{x})$. As $d_S(\bm{x}|\hat{\bm{x}}) = d_{RD}(\bm{x}|\hat{\bm{x}}) + \beta d_{T}(\bm{x}|\hat{\bm{x}})$ and $d_{T}(\bm{x}|\hat{\bm{x}})$ is related to $p(\bm{y}|\hat{\bm{x}})$ according to equation (\ref{eq_d_IB}). Therefore, the AI task performance will be influenced by the trade-off between pixel-level distortion and AI task-level distortion. When $p(\bm{y}|\hat{\bm{x}})$ is fixed,  we will go back to the rate distortion case with fixed distortion $d_{RD}(\bm{x},\hat{\bm{x}})$ and $d_{T}(\bm{x},\hat{\bm{x}})$. By iteratively solving the above self-consistent equations, we can obtain optimal mapping $p(\hat{\bm{x}}|\bm{x})$ and achieve the optimal trade-off between rate and semantic distortion.

\subsection{Discussion}

Above all, we derive a closed-form analytical solution for the extended rate-distortion optimization problem. Observing the self-consistent equations (\ref{eq_solution}), there are three challenges: 1) Accurate source probability $p(\bm{x})$ may be unavailable in practice; 2) The predicting model, $p(\bm{y}|\hat{\bm{x}})$, may not have analytical expression; 3) Even though there are accurate source probability $p(\bm{x})$ and analytical expression of $p(\bm{y}|\hat{\bm{x}})$, finding the optimal mapping $p(\hat{\bm{x}}|\bm{x})$ in the near-continuous search space may be impractical, especially for the high-dimensional data as images.

Therefore, to solve optimization problem (\ref{eq_optimization_2}), we adopt the deep learning method in Section IV. According to \cite{Balle, Theis, Toderici, ScaleHyperprior, CAECheng}, deep learning provides an excellent solution for seeking such a transformation from source to its representation. The essence of optimization problems (\ref{eq_optimization_2}) is to find the smallest semantic distortion of conditional probability $p(\hat{\bm{x}}|\bm{x})$ under the condition of certain compression. If the optimization goal is differential, the conditional probability $p(\hat{\bm{x}}|\bm{x})$ can be solved by using DNNs. In the next section we will build a feasible solution for TOSC-SR scheme based on DNNs.

\section{Proposed TOSC-SR Scheme}

Based on the theoretical analysis of the previous section, we will design a concrete TOSC-SR system. First, we introduce the components of the proposed system architecture in detail. Then, the practical loss function is given by relaxing the mutual information optimization objective for easy computation during the training process. Meanwhile, we point out that the loss function has close relationships with mutual information. Next, multi-task generalization validation is adopted to make the proposed system applicable to different AI tasks. Finally, corresponding training algorithms are presented.

\begin{figure*}[t]
\centering
\includegraphics[scale=0.7]{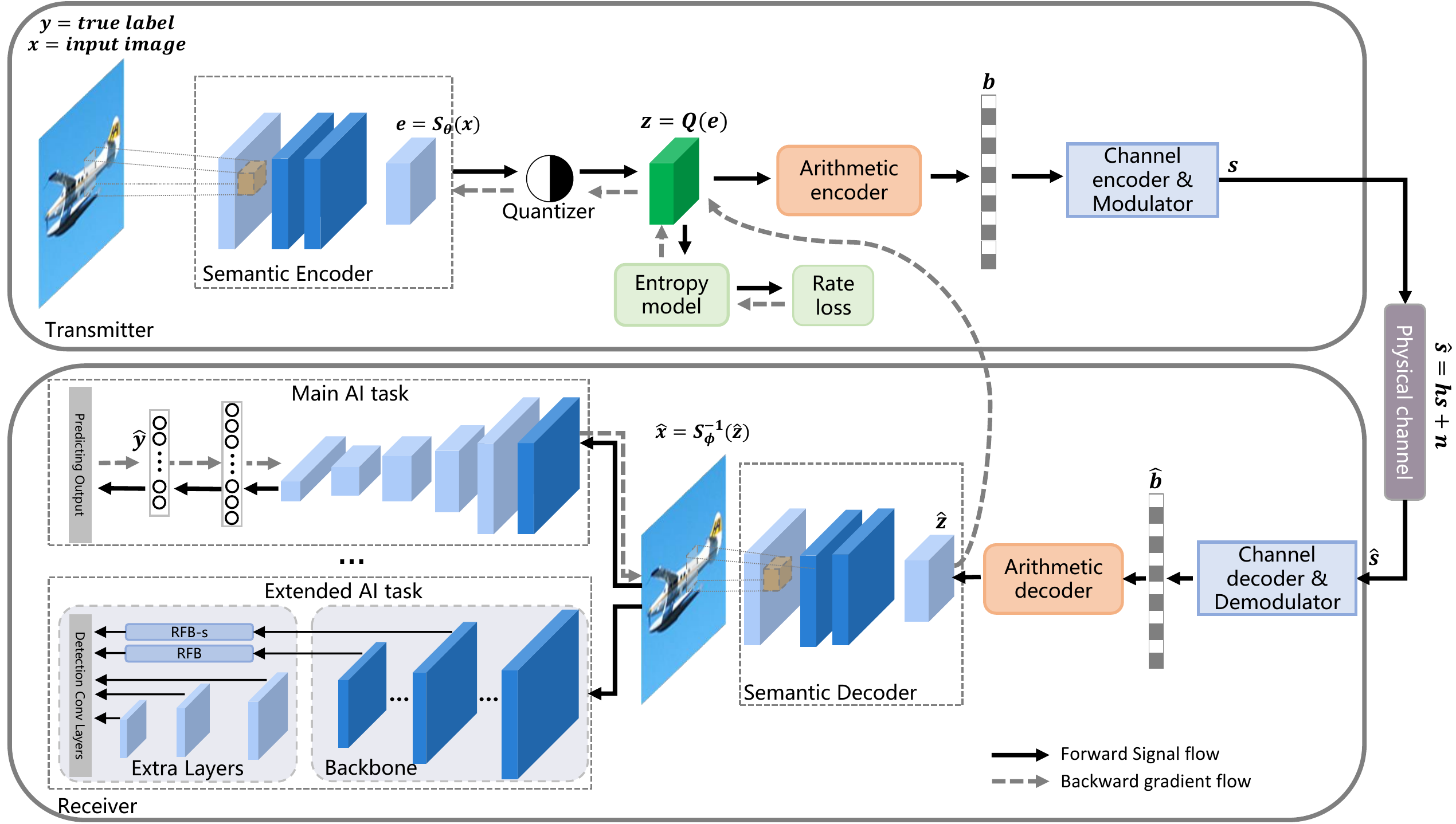}
\caption{The architecture of the TOSC-SR system.}
\label{fig_Semantic_image_transmission_overview}
\end{figure*}

\subsection{System Architecture}
The proposed system is shown in Fig. \ref{fig_Semantic_image_transmission_overview}. The overall architecture consists of three parts, i.e. transmitter, channel, and receiver. Transmitter is composed of a semantic encoder, a quantizer, an entropy model, an arithmetic encoder, a channel encoder and a modulator. Channel is the general Rayleigh fading  channel. Receiver is composed of a semantic decoder, an arithmetic decoder, a channel decoder, a demodulator, a main AI task, and an auxiliary AI tasks. Due to the non-differentiable nature, the arithmetic codec, channel codec, modulator, and demodulator are not involved in the training phase. In order to obtain the compression rate, entropy model is adopted at the transmitter. We adopt the ResNet-18 based classification network and RFB based object detection network \cite{RFBNet} as the main AI task and the auxiliary AI task, respectively. We use the classification task as our main AI task producing the training feedback on the semantic communication pipeline. Since the subsequent AI task of the receiver will have a positive impact on the performance of the whole network, the images reconstructed by the receiver contain more task-relevant information and have a certain generalization ability among different AI tasks. The implementation details of each part are described below.

At the transmitter, the semantic encoder consists of multiple convolution layers for dimensionality reduction and semantic extraction. Generalized divisive normalization (GDN), which has proven effective in Gaussianizing image densities \cite{GDN}, is adopted to introduce nonlinearity for image's semantic extraction. Quantization is necessary for further digital processing, while it is non-differentiable operation. The derivative of the rounding function is zero everywhere, rendering gradient descent ineffective. To enable the end-to-end differentiable training process, some smooth approximations are proposed in related works \cite{Balle, Theis, Toderici}. In the proposed method, differentiable rounding quantization is implemented by using a simple proxy function. In the forward propagation for training process, the quantizer is defined as
\begin{equation}\label{eq_quantization}
    \bm{z} = \bm{e} + \mathcal{U} \left (-\frac{1}{2}, \frac{1}{2} \right )
\end{equation}
where $\bm{e}$ is the semantic representation, $\mathcal{U} \left ( \cdot,\cdot \right )$ is the uniform noise vector. This method produces a proper approximation of regular rounding quantization. After quantization, the arithmetic encoder is used to compress the discrete semantic representation. In order to perform the entropy encoding better, we estimate the distribution of the quantized outputs, $\bm{z}$. We assume that the elements $z_i$ of $\bm{z}$ are i.i.d. with probability mass function (PMF) $p(z_i)$. Here, we formulate this PMF as a mixture of Gaussians. We first approximate $p(z_i)$ as a continuous-valued probability density funciton $p_c(z_i)$ as follows,
\begin{equation}
    p_c(z_i) = \sum_{m=1}^{M}\alpha_m \frac{1}{\sigma_m 2\pi}e^{-\frac{1}{2} \left ( \frac{z_i - \mu_m}{\sigma_m} \right )}
\end{equation}
where $M$ is the number of Gaussians mixtures, $\mu_m$ are mean values, $\sigma_m$ are scales and $\alpha_m$ are corresponding mixture weights. Then we integrate $p_c(z_i)$ over $\left [ z_i - \frac{1}{2}, z_i + \frac{1}{2} \right ]$ to obtain
\begin{equation}
    p(z_i) = \int_{z_i-\frac{1}{2} }^{z_i+\frac{1}{2}} {p_c(x)dx}.
\end{equation}
In order to obtain the parameters $\mu_m, \sigma_m $, and $\alpha_m$ of Gaussian mixture model, another sub-networks are designed to learn a probabilistic model over quantized semantic representation used for entropy coding. With the probability density function, we can easily estimate the entropy of semantic representation $z$, which can be used to train the overall end-to-end system. Here, we obtain the rate loss
\begin{equation}
    L_R = -\log_{2} p(\bm{z}).
\end{equation}
With this probability density function, we can also use it to help the arithmetic coding process in the inference phase. The channel coding and modulation are performed to alleviate the channel's affect. 

At the receiver, the noisy channel output symbols are first demodulated and decoded, obtaining the recovered bit sequence, $\hat{\bm{b}}$. The arithmetic decoder converts $\hat{\bm{b}}$ into semantic representation $\hat{\bm{z}}$ with the help of estimated probability density function of $\bm{z}$. The semantic decoder, which is almost symmetric to semantic encoder, performs the semantic reconstruction.

As the main AI task in the receiver, classification task plays a critical role in the TOSC-SR system. Without the classifier, this system would degrade to a normal codec for image transmission, which is optimized for the traditional rate-distortion performance. Once adding an AI task, which means introducing a task-relevant distortion, our codec is no longer an unsupervised network, the optimized objective will change from the previous rate-distortion trade-off $L = R+\lambda D_R$ to rate and semantic distortion trade-off $L_{S}=R+\lambda D_{R} +\beta D_{T}$. As shown in Fig. \ref{fig_Semantic_image_transmission_overview}, the classic ResNet-18 network followed by a fully connected layer is used as our classifier.

In order to validate the generalization ability of the proposed system on other different AI tasks, we choose the object detection as the validation task for the generalization ability of the proposed TOSC-SR scheme. A very important motivation to reconstruct images semantically is to ensure the generalization ability of the model between different tasks. For the multi-task learning model used in \cite{SPIC}, the network structure should be redesigned carefully if the receiver needs to perform a different task. In contrast, our method is highly flexible in transferring to other tasks. Specifically, we add a mature object detection network in parallel with classification network. The adopted object detection network is RFBNet\cite{RFBNet}. In the validation stage, the semantic reconstruction network trained with classification task remains unchanged.

\subsection{Loss Function and Mutual information}
In this section, we first derive the loss function used in the proposed TOSC-SR scheme. Due to the the complexity of mutual information computing, we derive a relaxed version of the loss function based on variational approximation\cite{VIB}. The relaxed loss function will be used for practical training. Since our loss function is based on the extended rate-distortion theory, we further analyze the relationship between the loss function and mutual information, which provides the information theory explanation to semantic reconstruction.

\subsubsection{Loss Function}
Following the optimization problem in (\ref{eq_optimization_2}), the goal of the system is to minimize the semantic distortion while reducing the rate, which results in the following loss function,
\begin{equation}\label{eq_loss_function_1}
    \mathcal{L} = R+\lambda D_{R}(\bm{X},\hat{\bm{X}}) + \beta D_{T}(\bm{X};\hat{\bm{X}}),
\end{equation}
where $R$ is the rate defined in (\ref{eq_rate}), $D_{R}(\bm{X},\hat{\bm{X}})$ is a fixed distortion function between $\bm{X}$ and $\hat{\bm{X}}$, i.e., MSE, $D_{T}(\bm{X};\hat{\bm{X}})$ is task-relevant distortion. As for the rate term,
\begin{equation}
    R = I(\bm{X};\hat{\bm{X}}) \le I(\bm{X};\bm{Z}) = H(\bm{Z}) - H(\bm{Z}|\bm{X}),
\end{equation}
where $H(\bm{Z}|\bm{X})=0$ since semantic encoder perform a deterministic transformation without uncertainty. Then we obtain the available upper limit of rate $R$, that is $H(\bm{Z})=-\log_2 p(\bm{z})$. As for the third term in (\ref{eq_loss_function_1}), minimizing $D_{T}(\bm{X}, \hat{\bm{X}})$ is equal to maximizing $I(\hat{\bm{X}};\bm{Y})$ according to Lemma \ref{lemma_1} since mutual information $I(\bm{X};\bm{Y})$ is a constant for a given dataset. While, $I(\hat{\bm{X}};\bm{Y})$ is still hard to compute.
According to the definition of mutual information
\begin{equation}
    I(\hat{\bm{X}};\bm{Y}) = H(\bm{Y})-H(\bm{Y}|\hat{\bm{X}}).
\end{equation}
Consider a variational approximation $q(\bm{y}|\hat{\bm{x}})$ of $p(\bm{y}|\hat{\bm{x}})$. The non-negativity of KL divergence leads to the following variational lower bound,
\begin{equation}
    \begin{aligned}
        I(\hat{\bm{X}};\bm{Y}) &= H(\bm{Y})-H(\bm{Y}|\hat{\bm{X}}) \\
                    &\geq H(\bm{Y}) - H(\bm{Y}|\hat{\bm{X}}) - D_{KL}(p(\bm{y}|\hat{\bm{x}})||q(\bm{y}|\hat{\bm{x}})) \\
                    &= H(\bm{Y}) + \mathbb{E}_{p(\bm{y},\hat{\bm{x}})}[\log q(\bm{y}|\hat{\bm{x}})],
    \end{aligned}
\end{equation}
where $p(\bm{y}|\hat{\bm{x}})$ is the distribution of a recognition model. And the second equality follows the following identity,
\begin{equation}
    \begin{aligned}
        -\mathbb{E}_{p(\bm{y},\hat{\bm{x}})}[\log q(\bm{y}|\hat{\bm{x}})] &= D_{KL}(p(\bm{y}|\hat{\bm{x}})||q(\bm{y}|\hat{\bm{x}})) + H(\bm{y}|\hat{\bm{x}}).
     \end{aligned}
\end{equation}
Then, we can have the following variational upper bound
\begin{equation}
    \begin{aligned}
        D_{T}(\bm{X},\hat{\bm{X}}) &= I(\bm{X};\bm{Y}) - I(\hat{\bm{X}};\bm{Y})\\
                          &\leq I(\bm{X};\bm{Y}) - H(\bm{Y}) - \mathbb{E}_{p(\bm{y},\hat{\bm{x}})}[\log q(\bm{y}|\hat{\bm{x}})].
     \end{aligned}
\end{equation}
Note that $I(\bm{X};\bm{Y}) - H(\bm{Y})$ is a constant, which is independent of the optimization process, and thus can be ignored. Then $\mathcal{L}$ can be further relaxed to minimize its upper bound, i.e.,
\begin{equation}
    \min H(\bm{Z}) + \lambda D_{R}(\bm{X},\hat{\bm{X}}) - \beta \mathbb{E}_{p(\bm{y},\hat{\bm{x}})}[\log q(\bm{y}|\hat{\bm{x}})].
\end{equation}
Note that $-\mathbb{E}_{p(\bm{y},\hat{\bm{x}})}[\log q(\bm{y}|\hat{\bm{x}})]$ is actually the cross-entropy loss, $CE(\bm{Y},\hat{\bm{Y}})$, in supervised learning. Above all, the adopted loss function in TOSC-SR is
\begin{equation} \label{eq_loss_function_2}
    \mathcal{L} = -\log_2 p(\bm{z}) + \lambda \mathbb{E}_{p(\bm{x},\hat{\bm{x}})}[||\bm{x}-\hat{\bm{x}}||^2] - \beta \mathbb{E}_{p(\bm{y},\hat{\bm{x}})}[\log q(\bm{y}|\hat{\bm{x}})].
\end{equation}

\subsubsection{Interpret Loss Function with Mutual Information}
In this section, we reveal the relationship between loss function and mutual information used in our method. Mutual information (MI) is a fundamental quantity for measuring the relationship between random variables. Minimizing or maximizing MI has gained considerable interests in a wide range of deep learning tasks. Similar to the idea in \cite{IB2015}, MI can be used as the measurement of the quality of DNN. In our proposed TOSC-SR scheme, MI also plays an important role in the following two aspects. First, we consider the information relevant to various target tasks as semantic information. MI provides a natural quantitative approach to relevant information. By encouraging the system to capture as much as meaningful information and discard as much as information irrelevant to any target tasks, we can realize a TOSC-SR system from the information theoretic view point. Second, the optimization objective applied in our system, i.e. (\ref{eq_loss_function_1}),  actually has close relationship with MI. The practical loss function adopted in the experiment, i.e. (\ref{eq_loss_function_2}), is a relaxed version of (\ref{eq_loss_function_1}). As shown in (\ref{eq_loss_function_1}), the optimization objective consists of rate $R$, reconstruction distortion $D_{RD}$, and single task relevant distortion $D_{T}$. Next we will prove that all of the three parts have close relationship with MI.

Rate $R = I(\bm{X};\hat{\bm{X}})$ is the MI between $\bm{X}$ and $\hat{\bm{X}}$. On the one hand, this MI determine the minimal number of bits per pixel, which is a theoretical lower bound. On the other hand, it also indicates the information about input $\bm{X}$ contained in code $\hat{\bm{X}}$. In general, the more this information is, the better reconstruction quality will be achieved. According to the data processing inequality,
\begin{equation}
    I(\bm{X};\hat{\bm{X}}) \leq I(\bm{X};\bm{Z}).
\end{equation}

$D_{RD}$ is the MSE between input $\bm{X}$ and reconstruction output $\hat{\bm{X}}$, as shown in equation (\ref{eq_MSE_loss_function}). MSE is the direct measurement of the difference between two variables while the MI quantify the similarity between two variables. These two metrics appear to be moving in opposite directions. In practice, $MSE(\bm{X};\hat{\bm{X}})$ is indeed related to $I(\bm{X};\bm{Z})$. According to the definition of MI, 
\begin{equation} \label{eq_MI_1}
    I(\bm{X};\bm{Z}) = H(\bm{X}) - H(\bm{X}|\bm{Z}),
\end{equation}
as $H(\bm{X})$ is a constant when dataset $\bm{X}$ is selected, we mainly focus on $H(\bm{X}|\bm{Z})$.
\begin{equation} \label{eq_MI_2}
    \begin{aligned}
         H(\bm{X}|\bm{Z}) &= -\sum_{\bm{x}}p(\bm{x})\sum_{\bm{z}}p(\bm{z}|\bm{x}) {\ln \left( p(\bm{x}|\bm{z}) \right)} \\
                &= \mathbb{E}_{\bm{x} \sim p(\bm{x})}\left[  \mathbb{E}_{\bm{z} \sim p(\bm{z}|\bm{x})} \left[ -\ln \left( p(\bm{x}|\bm{z}) \right) \right]  \right],
    \end{aligned}
\end{equation}
where $p(\bm{x}|\bm{z})$ is the semantic decoder. If we assume it follows the multivariate Gaussian distribution\begin{equation}
    \begin{aligned}
         p(\bm{x}|\bm{z})  &=  \mathcal{N}(\bm{x};\mu, \sigma^2 I_D),
    \end{aligned}
\end{equation}
where $\mu = D_{\phi}(\bm{z}) = \hat{\bm{x}}$. $D$ is the total dimension of $x$. So
\begin{equation}
    \begin{aligned}
         p(\bm{x}|\bm{z})  &=  \frac{1}{\prod_{i=1}^{D}{\sqrt{2 \pi \sigma_i ^2}}} \exp \left( -\frac{1}{2} \left \|  \frac{\bm{x}-\hat{\bm{x}}}{\sigma^2} \right \|^2 \right),
    \end{aligned}
\end{equation}
then,
\begin{equation} \label{eq_MI_3}
    \begin{aligned}
         -\ln \left( p(\bm{x}|\bm{z}) \right)  &= \frac{1}{2} \left \| \frac{\bm{x}-\hat{\bm{x}}}{\sigma^2} \right \|^2 + \frac{D}{2} \ln 2\pi + \frac{1}{2}\sum_{i=1}^{D}\ln \sigma_i ^2,
    \end{aligned}
\end{equation}
where the first term is the scaled MSE. Combing equations (\ref{eq_MI_1}), (\ref{eq_MI_2}) and (\ref{eq_MI_3}), we can find that if we use MSE as the loss function, minimizing $MSE(\bm{X};\hat{\bm{X}})$ means maximizing $I(\bm{X};\bm{Z})$. This is intuitive and reasonable that the more information about $\bm{X}$ is contained in $Z$, the better construction quality of $\hat{\bm{X}}$ is.

Single task relevant distortion $D_{T}$ is defined by the reduction of MI, as shown in (\ref{eq_DIB_2}). Minimizing $D_{T}$ means maximizing $I(\hat{\bm{X}};\bm{Y})$. According to the data processing inequality,
\begin{equation}
    I(\hat{\bm{X}};\bm{Y}) \geq I(\hat{\bm{Y}};\bm{Y}).
\end{equation}

Above all, $\bm{Z}$ and $\hat{\bm{Y}}$ are two critical variables in the whole end-to-end communication pipeline. Meanwhile, $\bm{Z}$ and $\hat{\bm{Y}}$ are bottleneck variables (which have the minimal number of dimension) of semantic codec and classifier, respectively. So computing MI about these two variables is necessary.


Despite excellent properties, the mutual information between high-dimensional variables has been quite difficult to calculate. Fortunately, there has emerged some sample-based MI estimation methods in recent years.

The mutual information neural estimator (MINE) in \cite{MINE} treats MI as the Kullback-Leibler (KL) divergence between the joint and marginal distributions and converts it into the dual representation as
\begin{equation}
    I_{\text{MINE}} := \mathbb{E}_{p(\bm{x},\bm{y})}\left[ T_\theta (\bm{x},\bm{y}) \right] - \log \left( \mathbb{E}_{p(\bm{x})p(\bm{y})}\left[ e^{T_\theta (\bm{x},\bm{y})} \right]    \right) ,
\end{equation}
where $T_\theta(\bm{x},\bm{y})$ is the function parametrized by a neural network, which takes samples $\bm{x}, \bm{y}$ as input.
Along with this method, the estimation of mutual information between high dimensional continuous random variables can be achieved by gradient descent over neural networks.

Recently, Cheng et al.\cite{CLUB} have introduced a contrastive log-ratio upper bound (CLUB).  Specifically, the CLUB bridges mutual information estimation with contrastive learning where MI is estimated by the difference of conditional probabilities between positive and negative sample pairs,
\begin{equation}
         I_{\text{CLUB}} := \mathbb{E}_{p(\bm{x},\bm{y})}\left[ \log p(\bm{y}|\bm{x}) \right] \mathbb{E}_{p(\bm{x})}\mathbb{E}_{p(\bm{y})}\left[ \log p(\bm{y}|\bm{x}) \right],
\end{equation}
where $p(\bm{y}|\bm{x})$ is the conditional distribution of $\bm{y}$ given $\bm{x}$. Without $p(\bm{y}|\bm{x})$, we can use a variational distribution $q(\bm{y}|\bm{x})$ to approximate $p(\bm{y}|\bm{x})$ and $q(\bm{y}|\bm{x})$ is usually implemented by a neural network.

MINE estimates the MI through the lower bound. CLUB estimates the MI through the upper bound. Regardless of upper or lower bound, we care more about the accuracy of estimation. According to the comparison experiments in \cite{CLUB}, CLUB achieves better accuracy.

\subsection{Training Algorithms}
In the proposed TOSC-SR scheme, the main training process consists of two phases: 1) pre-training the end-to-end image communication pipeline and image classifier separately, 2) jointly training the end-to-end image communication pipeline and image classifier. Since the channel coding and modulation are not DNN-based implementations, these two components are not included in the training phase. Algorithms \ref{alg_main}-\ref{algo_cae} describe the main training process. Algorithm \ref{algo_club} estimates the mutual information. Algorithm \ref{algo_transfer} is for validation of TOSC-SR's generalization ability.
\begin{algorithm}[h]
  \caption{TOSC-SR training algorithm.}
  \label{alg_main}
  \begin{algorithmic}[1]
    \Require The background knowledge $\mathcal{K}$, i.e. dataset.
    \State Train a classifier $F_{\chi}$ with input dataset $\mathcal{K}$;
    \State Train a end-to-end image semantic communication $S_{\bm{\theta}}$, $S^{-1}_{\bm{\phi}}$ with input dataset $\mathcal{K}$;
    \State Put the classifier behind the semantic codec, like Fig.\ref{fig_Semantic_image_transmission_overview}. Load all pre-trained network's parameters and freeze the parameters of $F_{\bm{\chi}}$;
    \While{Stop criterion is not met}
        \State \textbf{Forward:} $X\rightarrow \hat{\bm{X}} = S^{-1}_{\bm{\phi}}(S_{\bm{\theta}}(\bm{X}))$
        \State \qquad \qquad \ $\hat{\bm{X}} \rightarrow \hat{\bm{Y}}=F_{\bm{\chi}}(\hat{\bm{X}})$;
        \State \textbf{Loss:} $\mathcal{L}_R = MSE(\bm{X},\hat{\bm{X}}) + \beta CE(\bm{Y},\hat{\bm{Y}})$
        \State \textbf{Back-propagation:} $\mathcal{L}_R  \rightarrow \frac{\partial{\mathcal{L}_R}}{\partial{\bm{W_{cae}}}}$;
        \State \textbf{Update parameters:} $\bm{W_{cae}} := \bm{W_{cae}}- lr\frac{\partial{\mathcal{L}_R}}{\partial{\bm{W_{cae}}}}$
    \EndWhile
    \Ensure The parameterized networks $S_{\bm{\theta}}^{*}$ and $S^{-1*}_{\bm{\phi}}$.
  \end{algorithmic}  
\end{algorithm}

\begin{algorithm}[h]   
  \caption{Pre-train the classifier.} 
  \label{algo_cla}
  \begin{algorithmic}[1]  
    \Require The dataset $\mathcal{K}$.
    \State \textbf{Initialization:} Load the pre-trained feature extraction layer parameters of ResNet18 network, modify the output classes number of fully connection layer, then we can get the classification network's parameters $\bm{W_{cla}}$.
    \While{Stop criterion is not met}
        \State \textbf{Forward:} $\bm{X}\rightarrow \hat{\bm{Y}} = F_{\bm{\chi}}(\bm{X})$;
        \State \textbf{Loss:} $\mathcal{L}_{cla} = CE(\bm{Y}, \hat{\bm{Y}})$;
        \State \textbf{Back-propagation:} $\mathcal{L}_{cla} \rightarrow  \frac{\partial{\mathcal{L}_{cla}}}{\partial{\bm{W_{cla}}}}$;
        \State \textbf{Update parameters:} $\bm{W_{cla}} := \bm{W_{cla}}- lr\frac{\partial{\mathcal{L}_{cla}}}{\partial{\bm{W_{cla}}}}$
    \EndWhile
    \Ensure The parameterized network $F_{\bm{\chi}}$.
  \end{algorithmic}  
\end{algorithm}


\begin{algorithm}[h]   
  \caption{Pre-train the semantic codec.}  
  \label{algo_cae}
  \begin{algorithmic}[1]
    \Require The dataset $\mathcal{K}$.
    \State \textbf{Initialization:} Initialize the auto-encoder network's parameters $\bm{W_{cae}}$.
    \While{Stop criterion is not met}
        \State \textbf{Forward:} $\bm{X}\rightarrow \hat{\bm{X}} = S^{-1}_{\bm{\phi}}(S_{\theta}(\bm{X}))$;
        \State \textbf{Loss:} $\mathcal{L}_{cae} = MSE(\bm{X}, \hat{\bm{X}})$;
        \State \textbf{Back-propagation:} $\mathcal{L}_{cae} \rightarrow  \frac{\partial{\mathcal{L}_{cae}}}{\partial{\bm{W_{cae}}}}$;
        \State \textbf{Update parameters:} $\bm{W_{cae}} := \bm{W_{cae}}- lr\frac{\partial{\mathcal{L}_{cae}}}{\partial{\bm{W_{cae}}}}$
    \EndWhile
    \Ensure The parameterized network $S_{\bm{\theta}}$ and $S^{-1}_{\bm{\phi}}$.
  \end{algorithmic}  
\end{algorithm}

Algorithm \ref{alg_main} describes the training process of the proposed TOSC-SR scheme. In the preparation stage, a classifier and an convolutional auto-encoder need to be trained separately and then these two networks are incorporated into our communication system for joint fine-tuning training, as shown in Fig.1.

Algorithm \ref{algo_cla} and algorithm \ref{algo_cae} are subalgorithms of algorithm 1 and they are on how to train a classifier and an convolutional auto-encoder, respectively. The classifier is not trained from scratch because we only need to get a standard classifier here. With the help of classic ResNet18 network, pre-trained feature extraction layer parameters can be preloaded and then fine tuning training can be carried out to get a standard classification network. Auto-encoder is trained from scratch. See Algorithm 2 and Algorithm 3 for more details of these two training procedures. 

\begin{algorithm}[h]  
  \caption{Estimating Mutual Information $I(\bm{X};\bm{Z})$ Using CLUB method.}
  \label{algo_club}
  \begin{algorithmic}[1]
    \Require Dataset $\mathcal{K}$, fine-tuned auto-encoder $E_{\bm{\theta}}^{*}$, $D_{\bm{\phi}}^{*}$.
    \State \textbf{Initialization:} Initialize a variational network $q_{\bm{\xi}} (\bm{z}|\bm{x})$ with parameters $\bm{\xi}$.
    \State \textbf{Collect variables $\bm{X}$ and $\bm{Z}$:}  $\bm{X}\rightarrow \bm{Z} = E_{\bm{\theta}}(\bm{X})$ \quad $\bm{Z}\rightarrow \hat{\bm{X}} = D_{\bm{\phi}}(\bm{Z})$ \quad $\hat{\bm{X}} \rightarrow \hat{\bm{Y}}=T_{\bm{\psi}}(\hat{\bm{X}})$;
    
    \While{Stop criterion is not met}
        \State \textbf{Sampling:} Sample $\left\{ \left( x_i,y_i \right) \right\}_{i=1} ^N $ from $p(\bm{x},\bm{z})$;
        \State \textbf{Forward:} $X\rightarrow \mu,\sigma^2\rightarrow \hat{\bm{Z}}$;
        \State \textbf{Loss:} $\mathcal{L}_{\text{CLUB}} = MSE(\bm{Z}, \hat{\bm{Z}})$;
        \State \textbf{Back-propagation:} $\mathcal{L}_{\text{CLUB}} \rightarrow \frac{\partial{\mathcal{L}_{\text{CLUB}}}}{\partial{\bm{W_{\text{CLUB}}}}}$;
        \State \textbf{Update parameters:} $\bm{W_{\text{CLUB}}} := \bm{W_{\text{CLUB}}}- lr\frac{\partial{\mathcal{L}_{\text{CLUB}}}}{\partial{\bm{W_{\text{CLUB}}}}}$
    \EndWhile
    
    \State \textbf{Estimating MI:} $I_{\text{CLUB}} = \mathbb{E}_{p(\bm{x},\bm{y})}\left[ \log q_{\bm{\xi}}(\bm{y}|\bm{x}) \right]
         - \mathbb{E}_{p(\bm{x})}\mathbb{E}_{p(\bm{y})}\left[ \log q_{\bm{\xi}}(\bm{y}|\bm{x}) \right]$.
    \Ensure The estimating MI $I_{\text{CLUB}}$.
  \end{algorithmic}
  \label{transfer_algorithm}
\end{algorithm}

Algorithm \ref{algo_club} describes the steps of estimating mutual information provided by the CLUB method. As we have analyzed in the previous section, $\bm{Z}$ and $\hat{\bm{Y}}$ are bottleneck variables. Following the information bottleneck principle, mutual information $I(\bm{X};\bm{Z})$ and $I(\hat{\bm{Y}}; \bm{Y})$ are estimated using the CLUB method. Though different MIs need different estimating networks, these networks are similar to each other. We take $I(\bm{X};\bm{Z})$ as an example to exhibit the computing process in Algorithm \ref{algo_club}.

Algorithm \ref{algo_transfer} describes the specific steps of transfer from classification task to object detection task. Our goal is to achieve a trade-off between prediction precision and generalization ability. Prediction precision means that reconstructed images can better complete the task used in the training stage. Generalization ability means that if the task at the receiver changes, the reconstructed image can still have a better performance on the changed task. Here we take classification as the basic task and object detection as the target task. Note that What we do is a simple transfer, the only change is the task at the receiver, and we do not further train the semantic codec, which means that the network parameters used in the transfer experiments are the same as the fine tuning trained network with classification task.

\begin{algorithm}[h]
  \caption{Generalization ability validation: from classification to object detection.} 
  \label{algo_transfer}
  \begin{algorithmic}[1]
    \Require The dataset $\mathcal{K}$, fine-tuned auto-encoder $E_{\bm{\theta}}^{*}$, $D_{\bm{\phi}}^{*}$ and pre-trained object detection network $T_{\bm{\psi}}$.
    \State \textbf{Test with normal auto-encoder:} $\bm{X}\rightarrow \hat{\bm{X}} = D_{\bm{\phi}}(E_{\theta}(\bm{X}))$ \qquad $\hat{\bm{X}} \rightarrow \hat{\bm{Y}}=T_{\bm{\psi}}(\hat{\bm{X}})$;
    \State \textbf{Test with fine-tuned auto-encoder:} $\bm{X}\rightarrow \hat{\bm{X}}^{*} = D_{\bm{\phi}}^{*}(E_{\bm{\theta}}^{*}(\bm{X}))$ \qquad $\hat{\bm{X}}^{*} \rightarrow \hat{\bm{Y}}^{*}=T_{\bm{\psi}}(\hat{\bm{X}})$;
    \State Compare the mAP performance of $\hat{\bm{Y}} $ and $ \hat{\bm{Y}}^{*}$
    \Ensure The generalization ability.
  \end{algorithmic}
  \label{transfer_algorithm}
\end{algorithm}

\section{Experiments}
In this section, we provide extensive experiments to show the effectiveness of proposed task-oriented semantic communication with semantic reconstruction (TOSC-SR), which consists of three parts. First, we compare the proposed method with the traditional and the DNN-based methods in terms of the AI tasks performance and reconstruction quality under different bpp and SNR regimes. The visual results are also presented. Then, two specific mutual information are estimated to validate the assumption that our loss function has close relationship with mutual information. On the other hand, the changes of the these mutual information can also be used to explain the changes of performance in the first part of experiments. Finally, in order to validate the generalization among different AI tasks of the proposed TOSC-SR scheme, we test the generalization ability on object detection task. 

\subsection{Simulation settings}

\subsubsection{Datasets}
In these experiments, STL10 dataset \cite{STL_10} is adopted for classification and Pascal VOC 2007 dataset \cite{Pascal_VOC} is adopted for object detection. 

The STL-10 dataset is an image recognition dataset for developing deep learning algorithms. Specifically, there are 100,000 unlabeled images for unsupervised learning. The image size is $96 \times 96$. For supervised learning, it has 10 classes, each class has 500 training images and 800 test images with labels. Since our method is semi-supervised, we use the unlabeled images to train the end-to-end image communication, as shown in Algorithm \ref{algo_cae}, and utilize the labeled images to train the classifier.

The Pascal VOC 2007 dataset provides standardised image data sets for object class recognition. The dataset includes 9,963 images split into train/test sets, which include 5,011 and 4,952 images, respectively. We use our pre-trained image communication system to transmitting the test set and use the mature detection models to measure the mAP.

\subsubsection{Network and Hyper-parameters}

\begin{table}[t] \footnotesize
\centering
\caption{The Network Settings.}
\begin{tabular}{|c|c|c|c|}
\hline
\;      &Network type     &Layer      &Activation \\ 
\hline
\multirow{7}{*}{\tabincell{c}{Semantic \\ Encoder}} &Conv   &3xNx5x5, stride 2   &None   \\
\cline{2-4}     &GDN &\tabincell{c}{None} &None \\
\cline{2-4}     &Conv &NxNx5x5, stride 2 &None \\
\cline{2-4}     &GDN &\tabincell{c}{None} &None \\
\cline{2-4}     &Conv &NxNx5x5, stride 2 &None \\
\cline{2-4}     &GDN &\tabincell{c}{None} &None \\
\cline{2-4}     &Conv &NxMx5x5, stride 2 &None \\
\hline
\multirow{2}{*}{\tabincell{c}{Hyper \\ Encoder}}   &Conv &MxNx3x3, stride 1 &LeakyReLU \\
\cline{2-4}     &Conv &NxNx5x5, stride 2 &LeakyReLU \\
\hline
Quantizer   &\tabincell{c}{Uniform}   &Equation (\ref{eq_quantization})   &None   \\
\hline
Channel   &Rayleigh   &$\hat{z} = hz + n$   &None   \\
\hline
\multirow{2}{*}{\tabincell{c}{Hyper \\ Decoder}}   &Conv &NxMx5x5, stride 2 &LeakyReLU \\
\cline{2-4}     &Conv &Mx2Mx3x3, stride 1 &LeakyReLU \\
\hline
\multirow{7}{*}{\tabincell{c}{Semantic \\ Decoder}}   &Conv &MxNx5x5, stride 2 &None \\
\cline{2-4}     &GDN &\tabincell{c}{None} &None \\
\cline{2-4}     &Conv &NxNx5x5, stride 2 &None \\
\cline{2-4}     &GDN &\tabincell{c}{None} &None \\
\cline{2-4}     &Conv &NxNx5x5, stride 2 &None \\
\cline{2-4}     &GDN &\tabincell{c}{None} &None \\
\cline{2-4}     &Conv &Nx3x5x5, stride 2 &None \\
\hline
\tabincell{c}{Classification \\ Network}   &ResNet-18   &Default   &None   \\
\hline
\tabincell{c}{Object \\ Detection \\ Network}   &RFBNet   &Default   &None   \\
\hline
\multirow{4}{*}{MI Model}   &Linear1   &1024   &Relu   \\
\cline{2-4}     &Linear2 &512 &Relu \\
\cline{2-4}     &Linear3 &512 &Relu \\
\cline{2-4}     &Linear4 &Dimension of $z$  &None \\
\hline

\end{tabular}
\label{table_network_settings}
\end{table}

\begin{table}[t] \footnotesize
\centering
\caption{The Training Hyper-parameters Settings.}
\begin{tabular}{|c|c|c|c|c|}
\hline
\;      &Classifier &Semantic Codec     &TOSC-SR      &MI Model \\ 
\hline
\tabincell{c}{Initial \\ learning Rate} &1x$10^{-3}$   &\tabincell{c}{$10^{-4} \rightarrow 10^{-6}$}   &$1x10^{-6}$  &$1x10^{-4}$   \\
\hline
\tabincell{c}{Epochs} &1000   &1000   &200   &1000   \\
\hline
\tabincell{c}{Batch \\ Size} &32   &32   &32   &200   \\
\hline
\tabincell{c}{Optimizer} &SGD   &Adam   &Adam   &Adam   \\
\hline
\tabincell{c}{lr\_scheduler} &\tabincell{c}{reduce\_on\\ \_plateau}   &\tabincell{c}{reduce\_on\\ \_plateau}   &\tabincell{c}{reduce\_on\\ \_plateau}   &constant   \\
\hline
\end{tabular}
\label{table_parameters_settings}
\end{table}

The specific settings about various networks can be found in Table \ref{table_network_settings}. Note that the semantic encoder performs downsampling while implementing the convolutional operation by setting the stride larger than 1. In the semantic decoder, upsampling is performing by depth-to-space operation. In addition, The network structures of the classifier and objection detection are default settings of ResNet-18 and RFBNet, except that the fully connected layer in the ResNet-18 modified to 10 dimensions to adapt the practical task. Finally, the MI model is a variational network, which is to approximate the mappings between two variables.

The training parameters in Algorithms \ref{alg_main} to \ref{algo_club} are shown in Table \ref{table_parameters_settings}. Note that TOSC-SR is trained based on the separate classifier and semantic codec, so the learning rate is lower. The batch size of MI model is set to 200 since this model need rich data to reduce the estimation error. Trade-off factor $\beta$ and SNR only exist in our proposed TOSC-SR.

\subsubsection{Benchmarks}
The proposed TOSC-SR is essentially an image compression scheme for semantic communications. To implement a complete communication link, corresponding channel codecs, modulation schemes, and physical channel are required. In particular, the 2/3 rate LDPC, 16-QAM and AWGN channel with varying noise power $\sigma^2$ are adopted. There are two kinds of benchmarks. The first kind of benchmark adopts the handcrafted image compression schemes, such as JPEG, JPEG2000, BPG, VVC, followed by practical channel coding and modulation schemes over noisy channel. The second kind of benchmarks adopts the same network structure as the proposed TOSC-SR scheme, while the distortion functions are different. In particular, rate-reconstruction-distortion (RRD) adopts the MSE as its distortion function and rate-task-relevant-distortion (RTD) adopts the cross-entropy as its distortion function. Both RRD and RTD is followed by practical channel coding and modulation schemes over noisy channel. For the RRD, images are transmitted to minimize the reconstruction distortion and the received images are then input to the AI task's network. For the RTD, images are transmitted to minimize the task-relevant distortion and the received images are then input to the AI task's network.


\subsection{Comparison Experiments on STL-10}

\begin{figure*}[t]
\subfigure[PSNR.]
{
\begin{minipage}[t]{0.50\linewidth}
\includegraphics[width=3.22in]{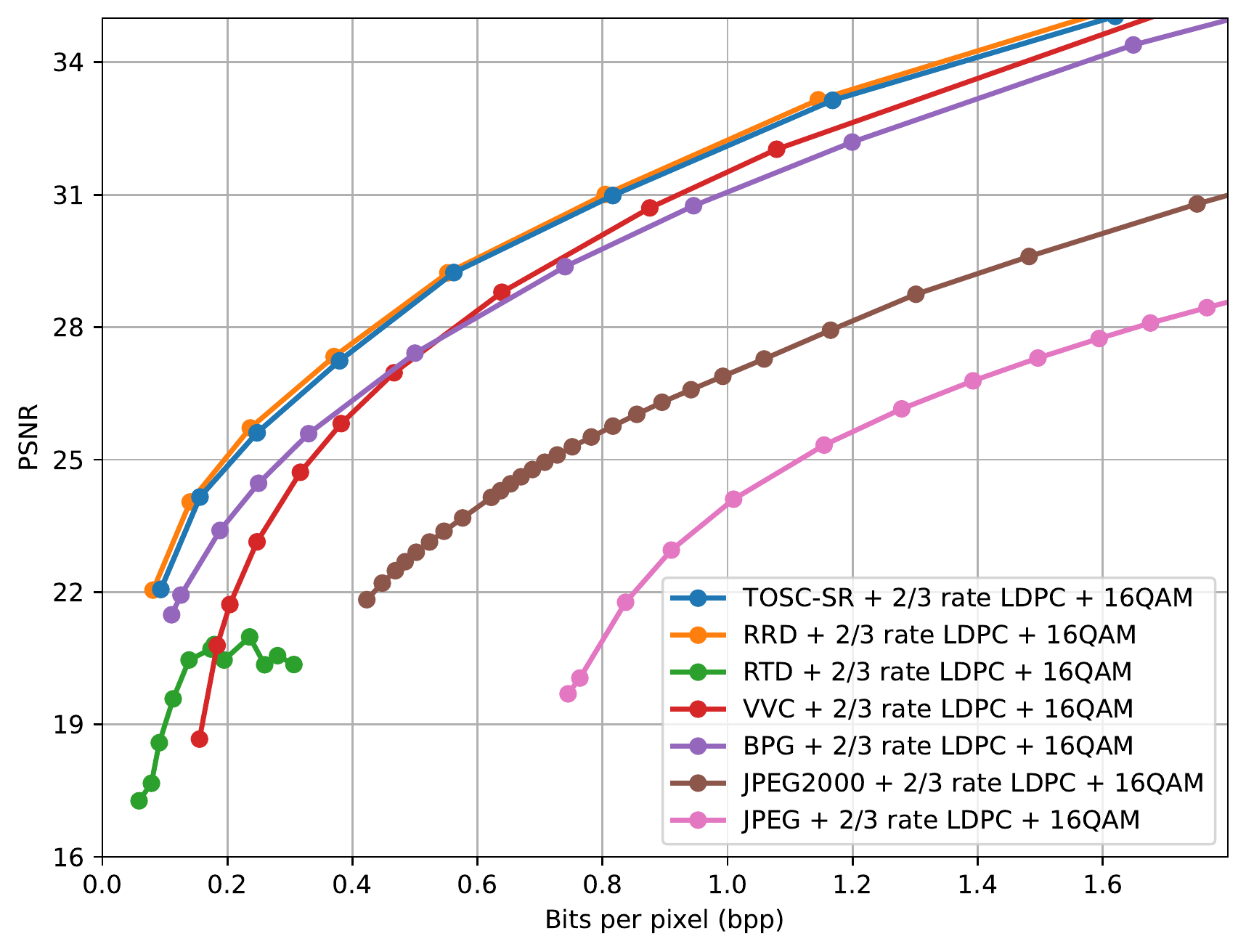}
\end{minipage}%
}%
\subfigure[Accuracy.]
{
\begin{minipage}[t]{0.50\linewidth}
\centering
\includegraphics[width=3.22in]{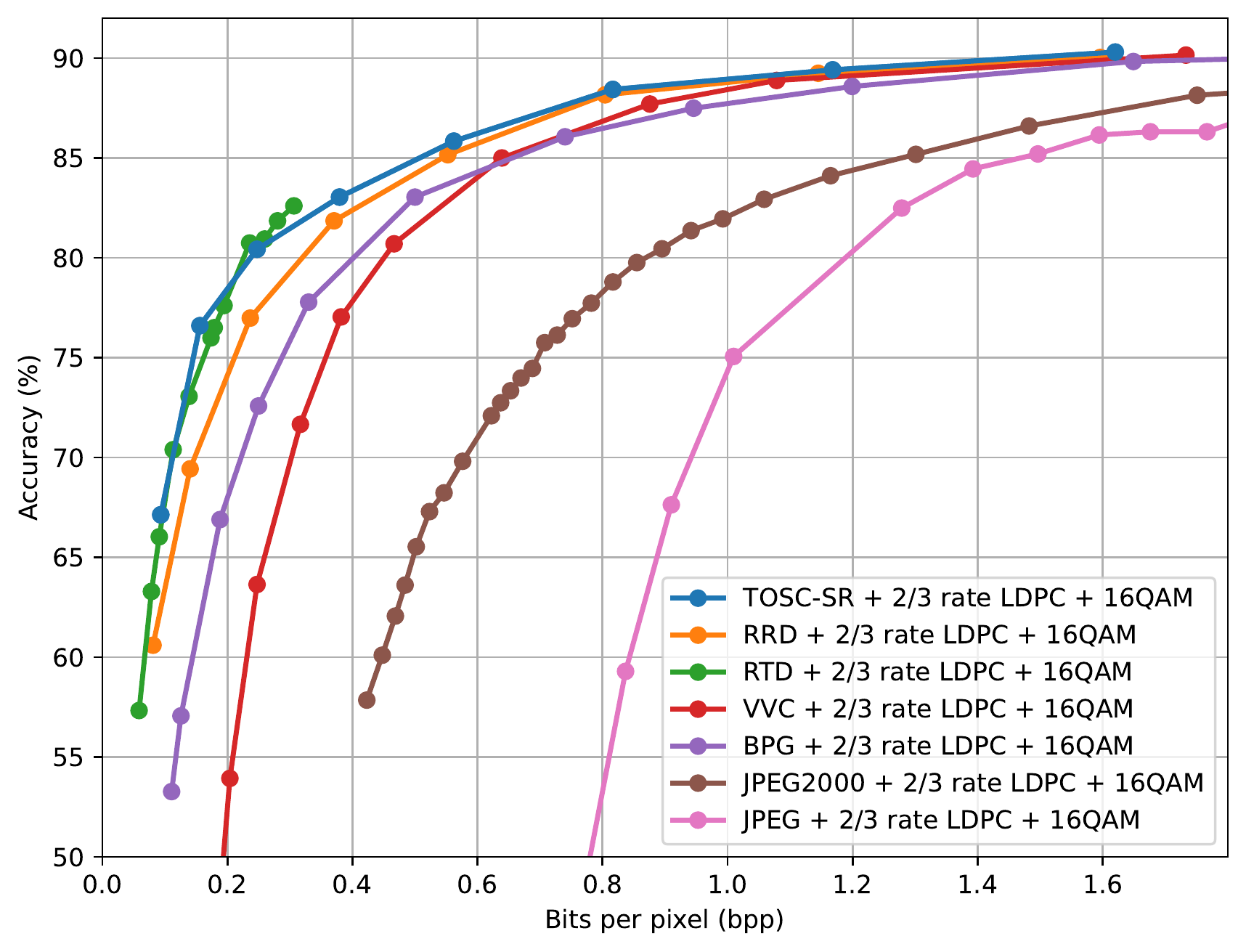}
\end{minipage}%
}%
\caption{Comparisons of different methods with respect to the performance of reconstruction and classification under different bpp with SNR = $20$, $\beta = 0.1$. The performance of reconstruction is assessed by PSNR and the performance of classification is assessed by classification accuracy.} 
\label{fig_performance_versus_bpp}
\end{figure*}

We use bits per pixel (bpp), peak signal-to-noise ratio (PSNR), and classification accuracy as the compression, reconstruction, and AI task's performance measurement. The classification accuracy for the original images is $91.36\%$. In the following analysis, we consider both the classification accuracy and reconstruction quality.

\subsubsection{Performance versus bpp}

Fig.\ref{fig_performance_versus_bpp} compares different methods with respect to the performance of reconstruction and classification under different bpp with SNR = $20$. For the TOSC-SR, $\beta = 0.1$. Fig.\ref{fig_performance_versus_bpp} (a) shows that TOSC-SR and RRD outperforms the handcrafted methods based schemes and RTD scheme in a large margin in terms of the reconstruction quality, which is consistent with the conclusions of most DNN-based image compression methods\cite{Balle, Theis, Toderici}. Moreover, TOSC-SR is competitive to RRD in terms of PSNR under all bpp settings. In addition, we find that RTD scheme show poor reconstruction ability due to the absence of MSE in its distortion funciton. Different from the reconstruction task, Fig.\ref{fig_performance_versus_bpp} (b) shows that the TOSC-SR and RTD outperform the other schemes in terms of classification accuracy under all bpp settings, especially in the low bpp. TOSC-SR is competitive to RTD in terms of classification accuracy under the lower bpp settings. It should be noted that RTD cannot achieve higher bpp results and the classification performance is nearly to converge, which limit the RTD's application in higher bpp settings. Overall, Fig.\ref{fig_performance_versus_bpp} shows that RTD attained the competitive  classification performance with TOSC-SR while the quality of reconstruction drops a lot compared with RRD and TOSC-SR, which will weaken the generalization ability among different tasks, i.e., these reconstructed images are unable to perform other tasks well. This will be further discussed in the later experiments. On the other hand, though the RRD baseline achieves the best quality of reconstruction, the classification performance is inferior to TOSC-SR, which supports our hypothesis that perceptional-significant visual features may not be the most suitable for AI task. The TOSC-SR achieved a ideal trade-off between predictive precision and generalization ability compared to the RRD by improving the reconstructed image's predictive precision in classification task at the expense of a small amount of perception quality.

\begin{figure*}[t]
\subfigure[PSNR.]
{
\begin{minipage}[t]{0.50\linewidth}
\includegraphics[width=3.22in]{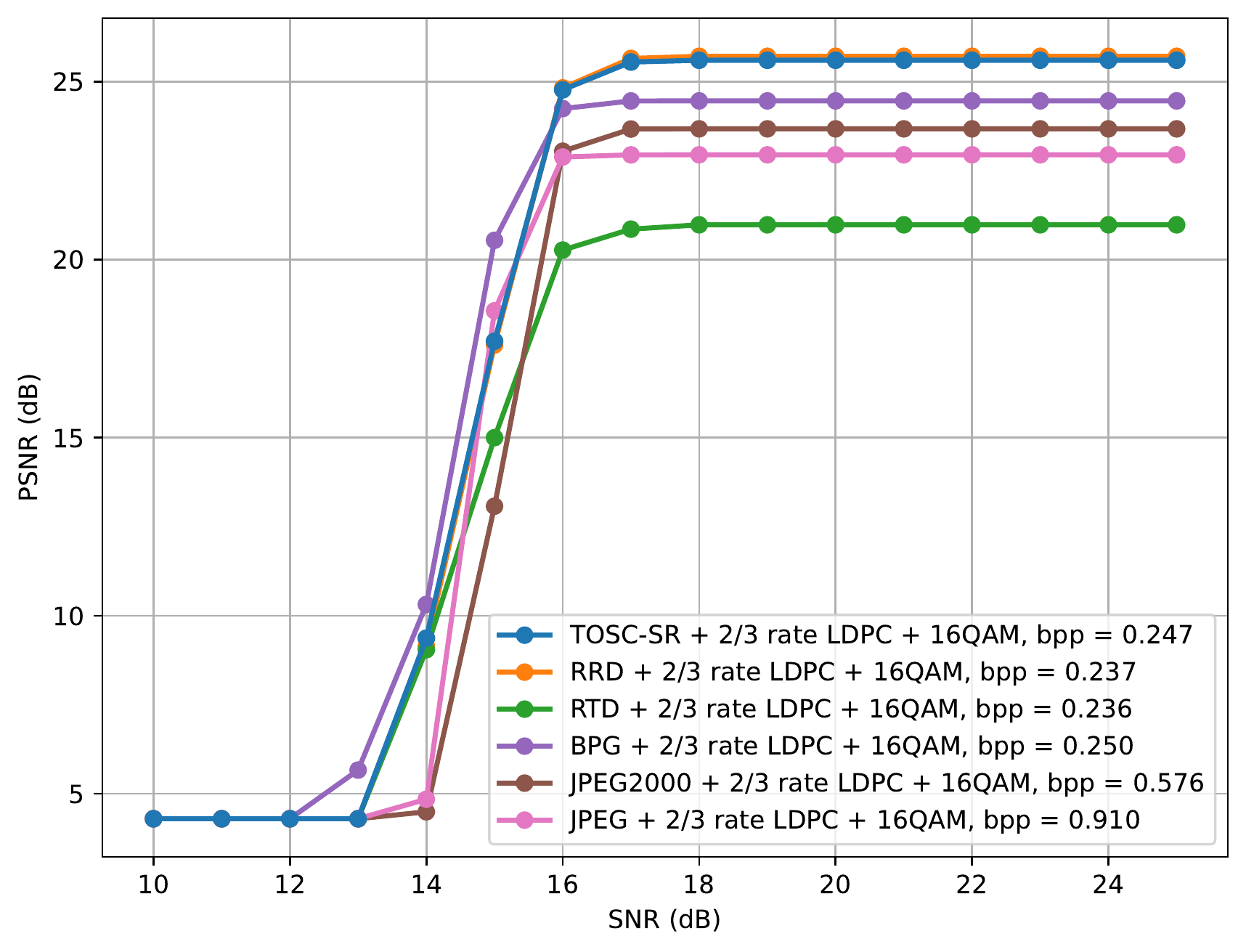}
\end{minipage}%
}%
\subfigure[Accuracy.]
{
\begin{minipage}[t]{0.50\linewidth}
\centering
\includegraphics[width=3.22in]{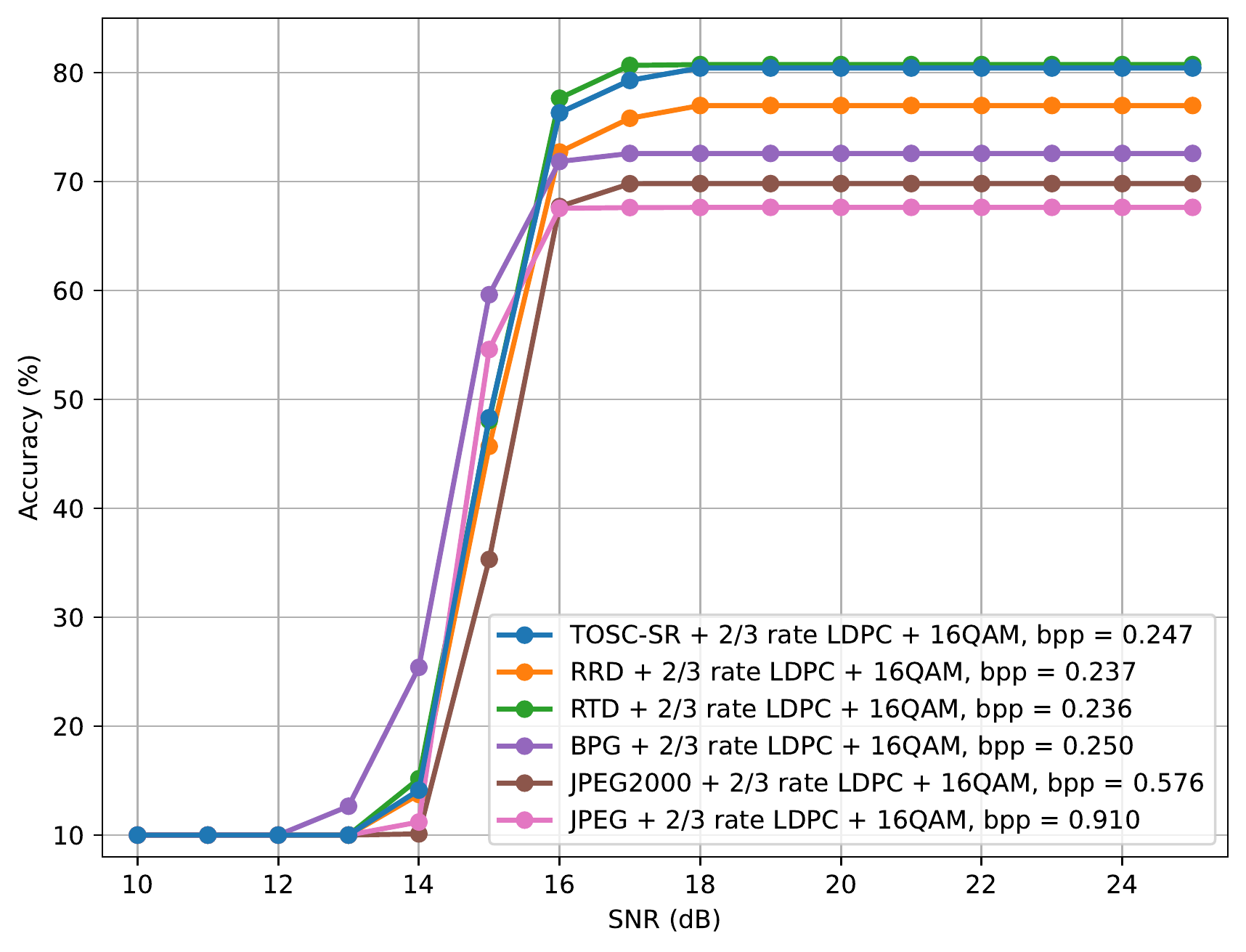}
\end{minipage}%
}%
\caption{Comparisons of different methods with respect to the performance of reconstruction and classification under different SNRs and similar bpps. For the TOSC-SR, $\beta = 0.1$. The performance of reconstruction is assessed by PSNR and the performance of classification is assessed by classification accuracy.} 
\label{fig_performance_versus_SNR}
\end{figure*}

\subsubsection{Performance versus SNR}
Fig.\ref{fig_performance_versus_SNR} compares the performance of reconstruction and classification versus SNRs. All the compared schemes use the same channel coding scheme, LDPC, and modulation scheme, 16-QAM. Due to the discrete nature of the compression ratio, we can only select approximated bpps for different methods. From Fig.\ref{fig_performance_versus_SNR} (a), we can observe that both TOSC-SR and RRD obviously outperform other schemes in terms of reconstruction performance, i.e. PSNR. Moreover, JPEG and JPEG2000 based schemes have the worst reconstruction performance even though their bpps are larger than other schemes. From Fig.\ref{fig_performance_versus_SNR} (b), we can observe that TOSC-SR and RTD can significantly outperform the other schemes in terms of classification accuracy. However, note that reconstruction performance of RTD is the worst as shown in \ref{fig_performance_versus_SNR} (a) since. In contrast, the proposed TOSC-SR can achieve both superior reconstruction and classification performance.

\subsubsection{Visualization of different schemes}
\begin{figure*}[t]
\centering
\includegraphics[scale=0.65]{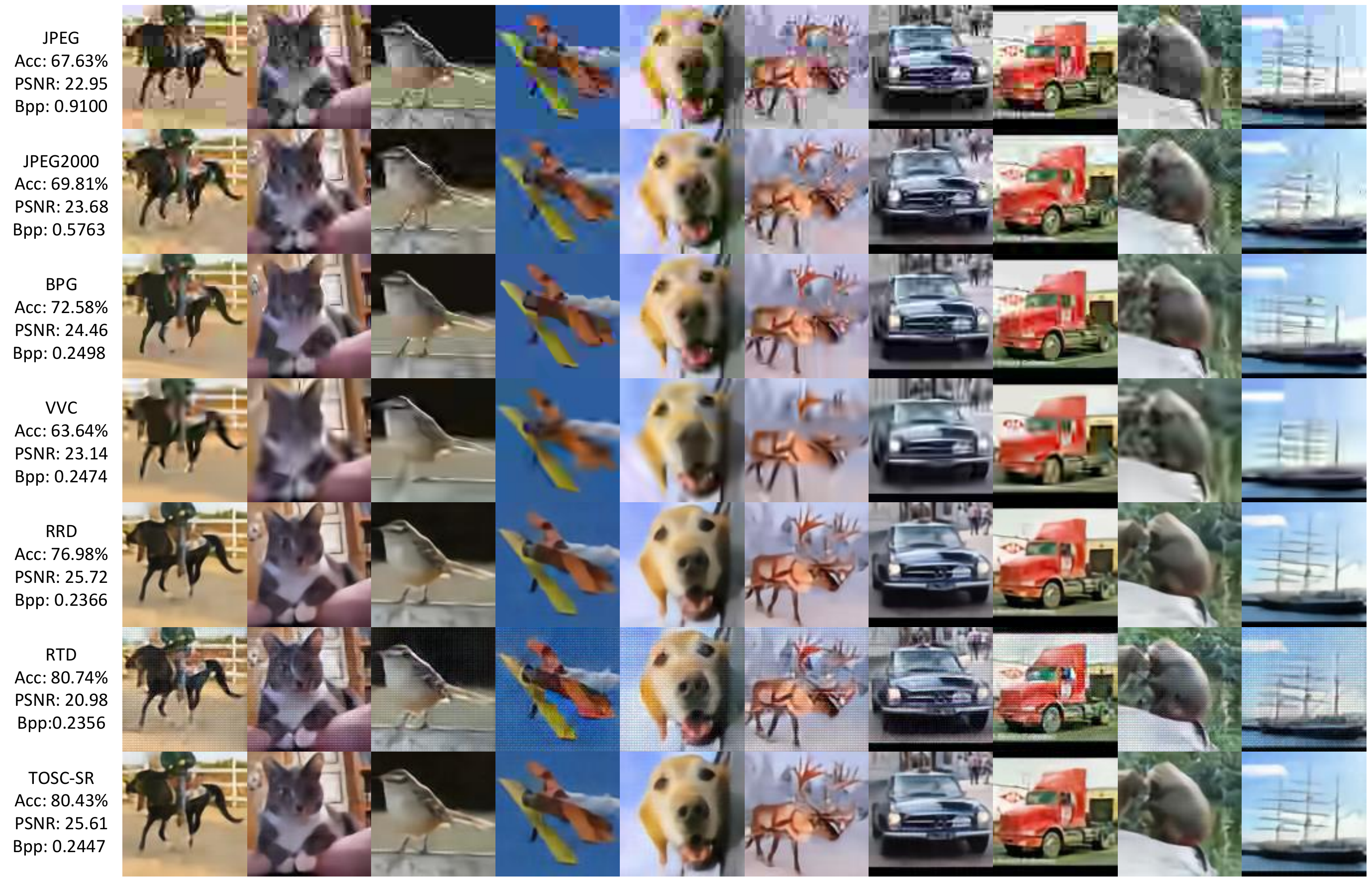}
\caption{Comparison of visual results of different schemes for STL-10 with SNR = $20$.}
\label{fig_classification_results_show}
\end{figure*}

Fig. \ref{fig_classification_results_show} shows the visualization results of reconstructed images using TOSC-SR scheme and the baseline schemes. From Fig.\ref{fig_classification_results_show}, we can observe that TOSC-SR and RRD outperform other schemes in terms of PSNR, while TOSC-SR and RTD achieve the highest classification accuracy. Therefore, in line with Fig.\ref{fig_performance_versus_SNR}, the visualization results also verify that the proposed TOSC-SR can achieve the best trade-off between reconstruction quality and AI task performance.




\subsection{Mutual Information Estimation Results}

\begin{figure*}[t]
    \subfigure[$I(\bm{X};\bm{Z})$.]{
    \begin{minipage}[t]{0.50\linewidth}
        \includegraphics[width=3.22in]{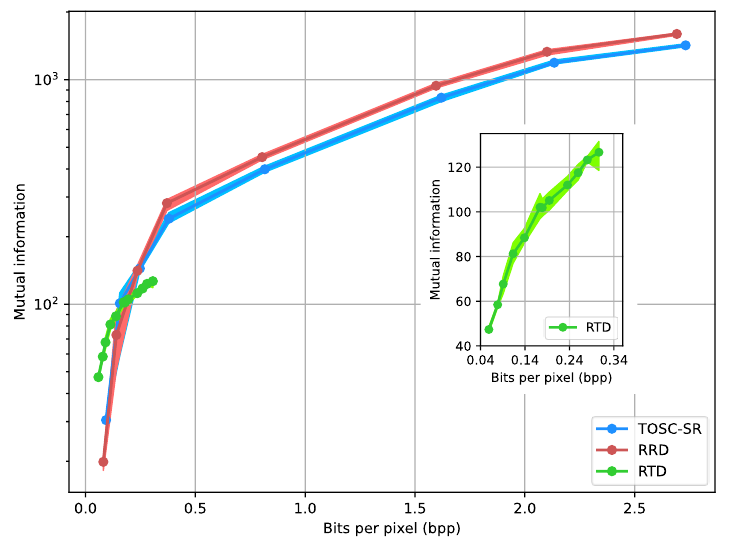}
    \end{minipage}%
    }%
    \subfigure[$I(\hat{\bm{Y}};\bm{Y})$.]{
        \begin{minipage}[t]{0.50\linewidth}
        \centering
        \includegraphics[width=3.22in]{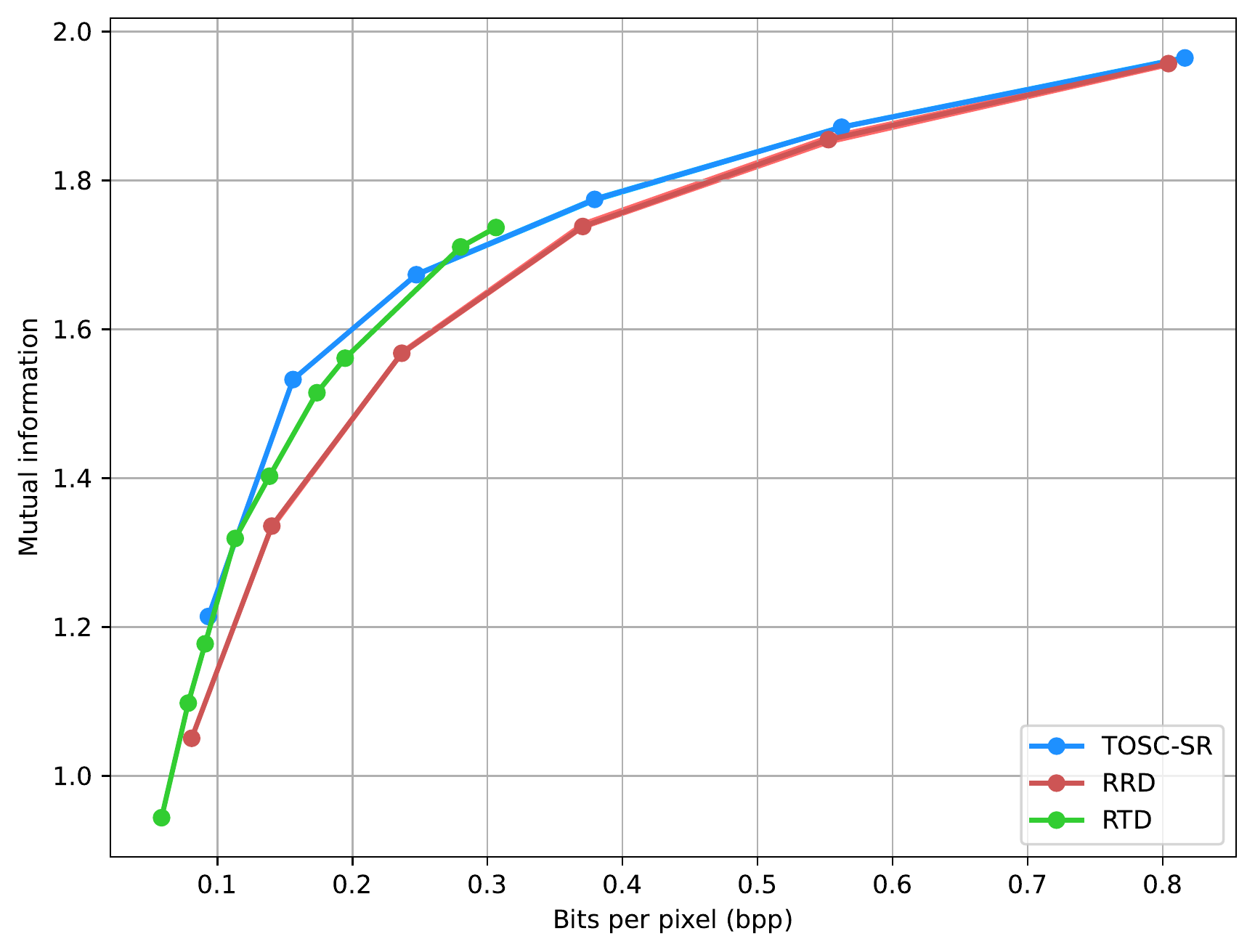}
    \end{minipage}
    }%
    \centering
    \caption{Comparisons of different methods with respect to four MI, $I(\bm{X};\bm{Z}), \text{and } I(\hat{\bm{Y}};\bm{Y})$ under different bpp with SNR = $20$, $\beta = 0.1$.}
    \label{fig_mutual_information}
\end{figure*}

\begin{figure*}[t]
\subfigure[PSNR]
{
\begin{minipage}[t]{0.50\linewidth}
\includegraphics[width=3.22in]{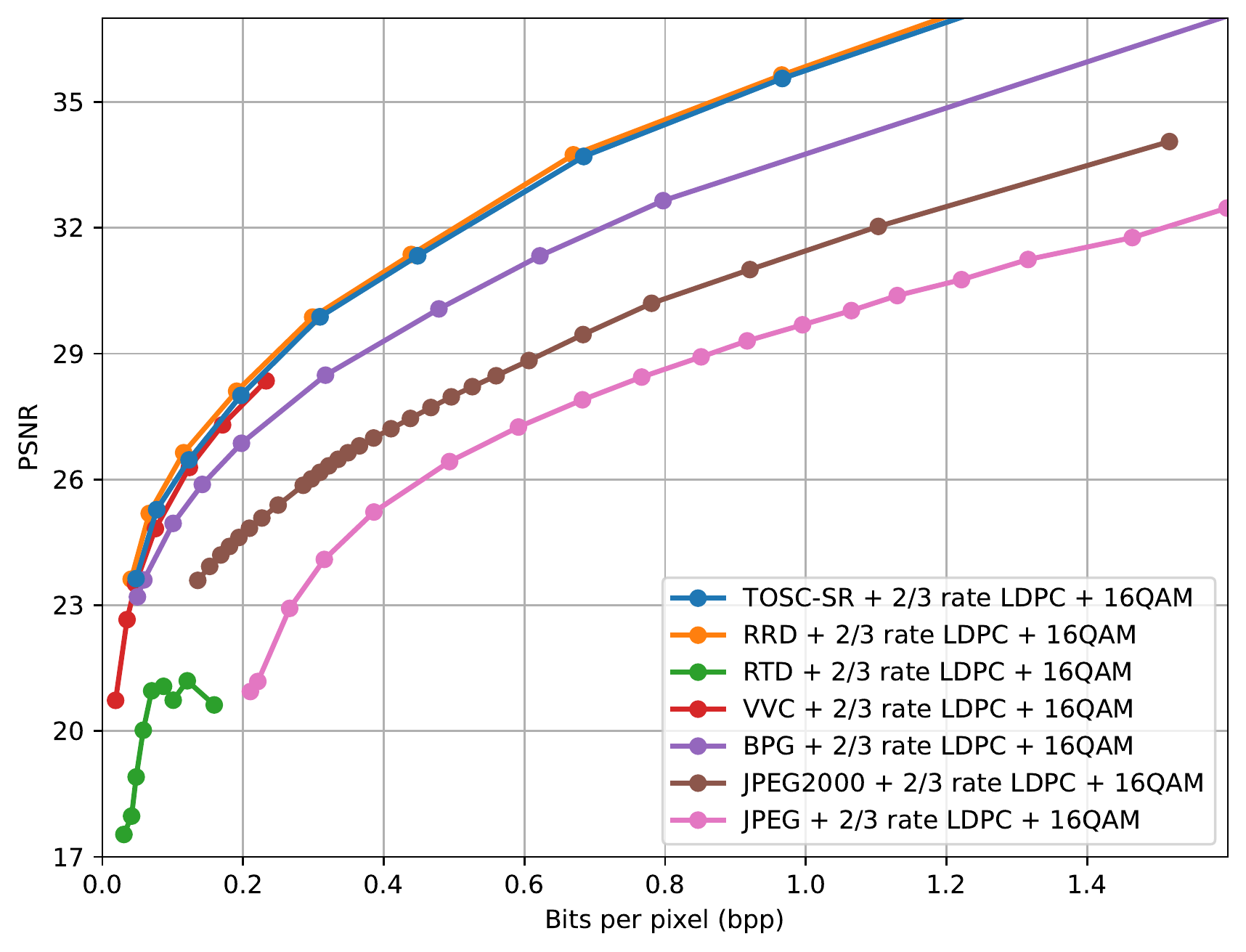}
\end{minipage}%
}%
\subfigure[mAP]
{
\begin{minipage}[t]{0.50\linewidth}
\centering
\includegraphics[width=3.22in]{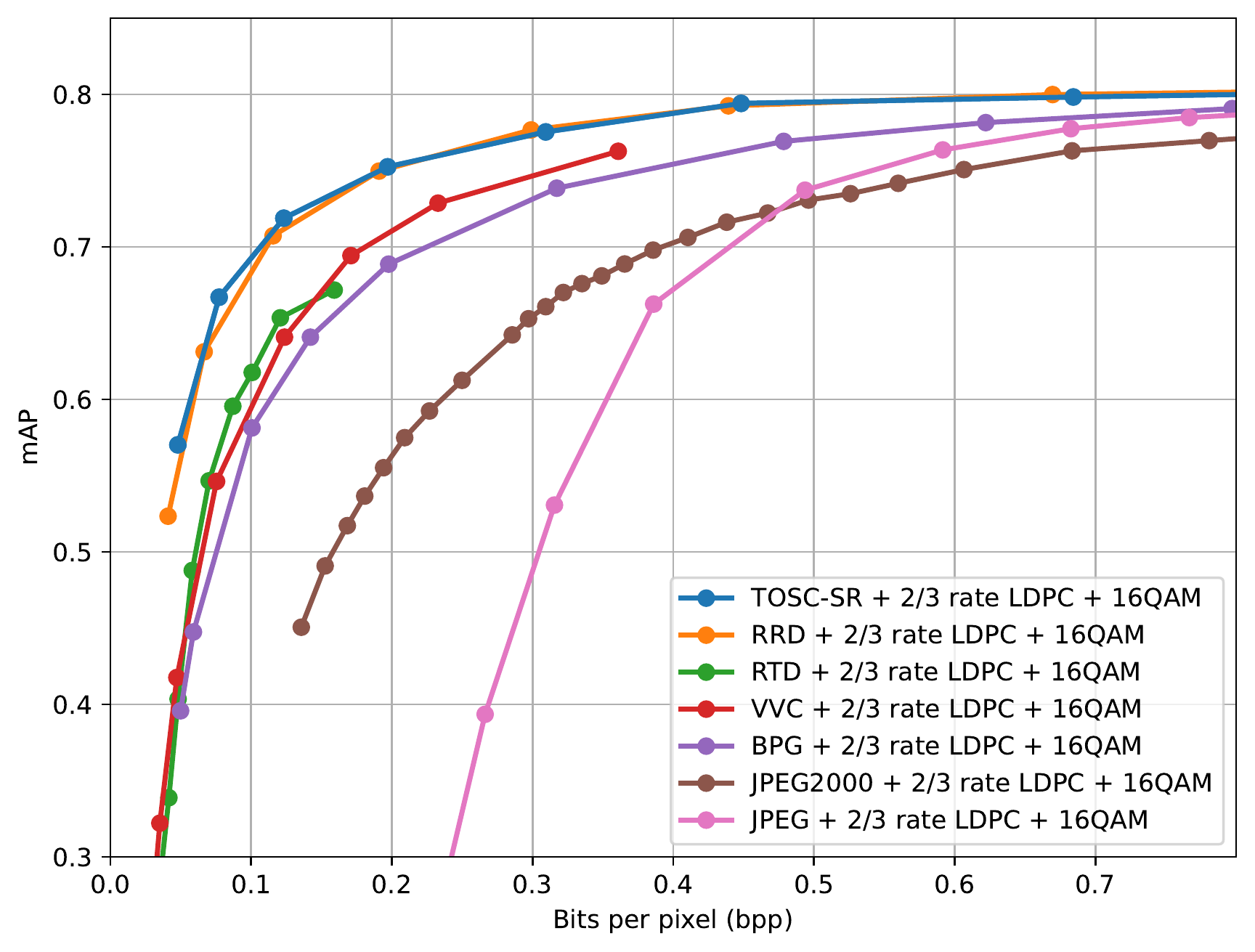}
\end{minipage}%
}%
\caption{Comparisons of different methods with respect to the performance of object detection and reconstruction under different bpp with SNR = $20$, $\beta = 0.1$. The performance of reconstruction is assessed by PSNR and the performance of object detection is assessed by mAP.}
\label{fig_voc_performance_versus_bpp}
\end{figure*}

Here, we will explain why the images with best visual perception are not necessarily the most suitable for the classification task from the perspective of information theory.

According to the previous analysis, random variable sets $\bm{Z}$ and $\hat{\bm{Y}}$ are two critical sets in the whole end-to-end communication pipeline. In this section we mainly estimate their MI values among TOSC-SR, RRD, and RTD. In line with information bottleneck method, we estimate two key mutual information $I(\bm{X};\bm{Z})$ and $I(\hat{\bm{Y}};\bm{Y})$. The physical meanings of these two MIs are described below. $I(\bm{X};\bm{Z})$ reflects the amount of information $\bm{Z}$ contains about $\bm{X}$. The larger this MI value, the better reconstruction potential the scheme will have. $I(\hat{\bm{Y}};\bm{Y})$ reflects the amount of information $\hat{\bm{Y}}$ contains about $\bm{Y}$. The larger this MI value, the better classification potential the scheme will have. Fig.(\ref{fig_mutual_information}) compares different methods in terms of two mutual information. It yields the following findings.
\begin{itemize}
    \item The loss functions used in different methods have close relationship with mutual information. From Fig.\ref{fig_mutual_information} (a), RRD, adopted the MSE-only objective, produces maximal $I(\bm{X};\bm{Z})$ since minimizing $MSE(\bm{X};\hat{\bm{X}})$ is equivalent to maximizing $I(\bm{X};\bm{Z})$. TOSC-SR, optimized for MSE and $D_{T}$ trade-off objective, exhibits a slightly smaller $I(\bm{X};\bm{Z})$ due to the influence of $D_{T}$. RTD shows the minimal $I(\bm{X};\bm{Z})$ because MSE is not included in the optimization process, and minimizing $D_{T}$ means maximizing $I(\hat{\bm{Y}};\bm{Y})$, which is different from minimizing $I(\bm{X};\bm{Z})$. Observing TOSC and RRD schemes in Fig.\ref{fig_mutual_information} (a) and Fig.\ref{fig_performance_versus_bpp} (a), we can find that they have similar dynamic curve. As for the RTD scheme, the curve in Fig.\ref{fig_mutual_information} (a) is not similar to Fig.\ref{fig_performance_versus_bpp} (a) due to the absence of MSE in the optimization process, which is explained before. On the contrary, from Fig.\ref{fig_mutual_information} (b), RRD produces the minimal $I(\hat{\bm{Y}};\bm{Y})$, RTD produces the maximal values and the values, of TOSC-SR is slightly lower than RTD. This phenomenon is also due to the different settings of optimization objectives.
    \item The change of mutual information results in the change of performance of reconstruction and classification. From Fig.\ref{fig_mutual_information} (a) and (b) and Fig.\ref{fig_performance_versus_bpp} (a) and (b), maximal $I(\bm{X};\bm{Z})$ and minimal $I(\hat{\bm{Y}};\bm{Y})$ of RRD correspond to the best reconstruction performance and worst classification performance. Minimal $I(\bm{X};\bm{Z})$ and maximal $I(\hat{\bm{Y}};\bm{Y})$ of RTD correspond to the worst reconstruction performance and best classification performance. Finally, the median MI values of TOSC-SR correspond to medium performance of different tasks. These results show that proper trade-off between different optimization objectives can obtain both satisfactory performance of various tasks.
\end{itemize}

If the distorted description is merely $D_{RD}$, which is actually MSE here, then this pixel-by-pixel approach is equivalent to preserving as much feature information as possible of the original image without discriminating and any trade-off. However, we believe that in the case of limited bit rate, it is not beneficial to the execution of downstream AI tasks if the image is still recovered with MSE as the target. Specifically, the network is no longer able to recover the original image well in the low bit rate. If we still try to recover the image pixel-by-pixel, it will become less recognizable.

If the distorted description is only $D_{T}$, recovered $\hat{\bm{X}}$ will retain as much as possible of the feature information relevant to corresponding task. Although this is good for the task, the structural information of the original image is completely destroyed, making it impossible to perform other tasks, especially those tasks need the structural information. In fact, it is almost impossible to reconstruct the a meaningful image for human eyes.

According to our method, the loss of two kinds of distortions, MSE and mutual information $I(\hat{\bm{X}};\bm{Y})$, are jointly considered, so that the task-related feature information can be selectively recovered and the generalization ability can be retained. In the case of limited bit rate, it not only ensures the performance of completing the task but also enables the network to reconstruct the image with slightly reduced quality. At this time, the reconstructed image has already contained more feature information related to the task, which is called semantic reconstruction.

\subsection{Generalization Performance on Pascal VOC}

Fig. \ref{fig_voc_performance_versus_bpp} shows the generalization performance of different methods from classification to object detection. Note that RFBNet's mAP on the original Pascal VOC 2007 test dataset is $80.67\%$. From Fig. \ref{fig_voc_performance_versus_bpp} (a), similar the results of STL-10 dataset, TOSC-SR and RRD outperform other schemes in terms of PSNR under all bpp settings. While, unlike the classification case, TOSC-SR and RRD outperform RTD and the traditional benchmarks in a large margin in terms of mAP in Fig. \ref{fig_voc_performance_versus_bpp} (b). Though RTD obtains the competitive performance with TOSC-SR in the classification task, RTD cannot maintain its advantage in the object detection task, which verify the good generalization ability among different AI tasks of the proposed TOSC-SR scheme. For RTD scheme, the quality of its reconstruction drops a lot, which is not suitable for the AI tasks except for classification.

\section{Conclusion}
In this paper, we proposed a TOSC-SR system based on the extended rate-distortion, which is jointly optimized for the predicting precision and generalization ability among different AI tasks of the reconstructed images. Specifically, we extended the rate-distortion theory by taking the trade-off between reconstruction distortion and task-related distortion as the semantic distortion and further designed the semantic communication system based on the widely used semantic codec architecture followed by the task network. The semantic distortion we define here is closely related to mutual information. Therefore our semantic communication system can be explained from the information theoretic perspective. The simulation results have shown that the proposed method outperforms the traditional and DNN-based methods in both classification and object detection tasks under different bpp and SNR. The mutual information estimation results show that our TOSC-SR indeed extracts more semantic information. Therefore, our proposed method can be used in IoT communications to perform various AI tasks.

\begin{appendices}
      \section{Proof of Theorem 1}
      \label{App_A}
      Denote $d_{T}(\bm{x},\hat{\bm{x}})$ as the relevant information distortion between $\bm{x}$ and $\hat{\bm{x}}$,
      \begin{equation}
          d_{T}(\bm{x},\hat{\bm{x}}) = \sum_{\bm{y}} p(\bm{y}|\bm{x})\log \left(\frac{p(\bm{y}|\bm{x})}{p(\bm{y}|\hat{\bm{x}})} \right).
      \end{equation}
      Denote $D_{T}(\bm{x},\hat{\bm{x}})$ as the expectation of relevant information distortion between $\bm{x}$ and $\hat{\bm{x}}$,
      \begin{equation}
        \begin{aligned}
             D_{T}(\bm{x},\hat{\bm{x}}) &= E[d_{T}(\bm{x},\hat{\bm{x}})] \\
             &= \sum_{\hat{\bm{x}}}\sum_{\bm{x}}\sum_{\bm{y}}p\left(\bm{x},\hat{\bm{x}}\right)p\left(\bm{y}\middle|\bm{x}\right)\ \log{\left(\frac{p\left(\bm{y}\middle|\bm{x}\right)}{p\left(\bm{y}\middle|\hat{\bm{x}}\right)}\right)}.
        \end{aligned}
      \end{equation}
      We first prove conditional mutual information $I(\bm{X};\bm{Y}|\hat{\bm{X}})$ does describe the relevant information distortion. According to the markov chain $\bm{Y} \rightarrow \bm{X} \rightarrow \hat{\bm{X}}$,
      \begin{equation}\label{eq_markov}
          p(\bm{y}|\bm{x},\hat{\bm{x}})=p(\bm{y}|\bm{x}).
      \end{equation}
      According to the definition of conditional mutual information, and implement equation (\ref{eq_markov})
      \begin{equation}
        \begin{aligned}
            &\phantom{\;=} I\left(\bm{X};\bm{Y}\middle|\hat{\bm{X}}\right) \\
            &= \sum_{\hat{\bm{x}}}\sum_{\bm{x}}\sum_{\bm{y}}p\left(\hat{\bm{x}}\right)p\left(\bm{x},\bm{y}\middle|\hat{\bm{x}}\right)\log{\left(\frac{p\left(\bm{x},\bm{y}\middle|\hat{\bm{x}}\right)}{p\left(\bm{x}\middle|\hat{\bm{x}}\right)p\left(\bm{y}\middle|\hat{\bm{x}}\right)}\right)}\\
            &= \sum_{\hat{\bm{x}}}\sum_{\bm{x}}\sum_{\bm{y}}p\left(\hat{\bm{x}}\right)p\left(\bm{x},\bm{y}\middle|\hat{\bm{x}}\right)\log{\left(\frac{p\left(\bm{y}\middle|\bm{x},\hat{\bm{x}}\right)p(\bm{x}|\hat{\bm{x}})}{p\left(\bm{x}\middle|\hat{\bm{x}}\right)p\left(\bm{y}\middle|\hat{\bm{x}}\right)}\right)}\\
            &= \sum_{\hat{\bm{x}}}\sum_{\bm{x}}\sum_{\bm{y}}p\left(\hat{\bm{x}}\right)p\left(\bm{y}\middle|\bm{x},\hat{\bm{x}}\right)p(\bm{x}|\hat{\bm{x}})\log{\left(\frac{p\left(\bm{y}\middle|\bm{x}\right)}{p\left(\bm{y}\middle|\hat{\bm{x}}\right)}\right)}\\
            &= \sum_{\hat{\bm{x}}}\sum_{\bm{x}}\sum_{\bm{y}}p\left(\bm{x},\hat{\bm{x}}\right)p\left(\bm{y}\middle|\bm{x}\right)\log{\left(\frac{p\left(\bm{y}\middle|\bm{x}\right)}{p\left(\bm{y}\middle|\hat{\bm{x}}\right)}\right)}\\
            &= D_{T}=E[d_{T}(\bm{x},\hat{\bm{x}})].\\
        \end{aligned}
      \end{equation}
      
      Then we prove
      \begin{equation}
          I(\bm{X};\bm{Y})-I(\hat{\bm{X}};\bm{Y}) == I(\bm{X};\bm{Y}|\hat{\bm{X}}),
      \end{equation}
      
      According to the definition of mutual information,
      \begin{equation}
          \begin{aligned}
            &\phantom{\;=} I(\bm{X};\bm{Y})-I(\hat{\bm{X}},Y)\\
            &= \sum_{\bm{x}} \sum_{\bm{y}} {p(\bm{x},\bm{y})\log \left(\frac{p(\bm{x},\bm{y})}{p(\bm{x})p(\bm{y})} \right)} \\
            &\phantom{\;=} - \sum_{\hat{\bm{x}}} \sum_{\bm{y}} {p(\hat{\bm{x}},\bm{y})\log \left(\frac{p(\hat{\bm{x}},\bm{y})}{p(\hat{\bm{x}})p(\bm{y})} \right)} \\
            &= \sum_{\hat{\bm{x}}}\sum_{\bm{x}} \sum_{\bm{y}} {p(\hat{\bm{x}},\bm{x},\bm{y})\log \left(\frac{p(\bm{x},\bm{y})}{p(\bm{x})p(\bm{y})} \right)} \\ &\phantom{\;=} - \sum_{\hat{\bm{x}}} \sum_{\bm{x}} \sum_{\bm{y}} {p(\hat{\bm{x}},\bm{x},\bm{y})\log \left(\frac{p(\hat{\bm{x}},\bm{y})}{p(\hat{\bm{x}})p(\bm{y})} \right)} \\
            &= \sum_{\hat{\bm{x}}}\sum_{\bm{x}} \sum_{\bm{y}} {p(\hat{\bm{x}},\bm{x},\bm{y})\log \left(\frac{p(\bm{y}|\bm{x})}{p(\bm{y})} \right)} \\
            &\phantom{\;=} - \sum_{\hat{\bm{x}}} \sum_{\bm{x}} \sum_{\bm{y}} {p(\hat{\bm{x}},\bm{x},\bm{y})\log \left(\frac{p(\bm{y}|\hat{\bm{x}})}{p(\bm{y})} \right)} \\
            &= \sum_{\hat{\bm{x}}}\sum_{\bm{x}} \sum_{\bm{y}} {p(\hat{\bm{x}},\bm{x},\bm{y})\log \left(\frac{p(\bm{y}|\bm{x})}{p(\bm{y}|\hat{\bm{x}})} \right)}\\
            &= D_{T}=E[d_{T}(\bm{x},\hat{\bm{x}})].
          \end{aligned}
      \end{equation}
      
      \section{Proof of Theorem 2}
      \label{App_B} 
      Rewrite the optimization problem (\ref{eq_optimization_2}) as follows:
      \begin{equation}
        \underset{
            \substack{
            p(\hat{\bm{X}}|\bm{X}):
            I(\bm{X},\hat{\bm{X}}) \leq I_C\\
            \sum_{\hat{\bm{x}}}{p(\hat{\bm{x}}|\bm{x})} = 1
            }
        }
        {\min} D_{RD}(\bm{X},\hat{\bm{X}}) - \beta I(\hat{\bm{X}};\bm{Y}).
      \end{equation}
      
        Using the Lagrange multiplier method by first formulating the Lagrange functional,
        \begin{equation}
            \begin{aligned}
                \mathcal{L}\left(p\left(\hat{\bm{x}}|\bm{x}\right)\right) &= \lambda I(X;\hat{\bm{X}}) +  D_{RD}(X;\hat{\bm{X}}) - \beta I(\hat{\bm{X}};\bm{Y}) \\
                &\phantom{\;=} + \sum_{\bm{x}}r(\bm{x})\sum_{\hat{\bm{x}}}{p(\hat{\bm{x}}|\bm{x})} \\
                &=\lambda \sum_{\bm{x}}\sum_{\hat{\bm{x}}}{p(\bm{x})p(\hat{\bm{x}}|\bm{x}) \log \frac{p(\hat{\bm{x}},\bm{x})}{p(\hat{\bm{x}})}} \\
                &\phantom{\;=} + \sum_{\bm{x}}\sum_{\hat{\bm{x}}}{p(\bm{x})p(\hat{\bm{x}}|\bm{x})d_{RD}(\bm{x},\hat{\bm{x}})} \\
                &\phantom{\;=} - \beta \sum_{\hat{\bm{x}}} \sum_{\bm{y}}p(\bm{y})p(\hat{\bm{x}}|\bm{y})\log \frac{p(\hat{\bm{x}}|\bm{y})}{p(\hat{\bm{x}})} \\
                &\phantom{\;=} + \sum_{\bm{x}}r(\bm{x})\sum_{\hat{\bm{x}}}{p(\hat{\bm{x}}|\bm{x})},
            \end{aligned}
        \end{equation}
        where $\lambda$ is the Lagrange multiplier attached to the constrained $I(X;\hat{\bm{X}})$, $r(\bm{x})$ is the Lagrange multiplier attached to the normalization of the mapping $p(\hat{\bm{x}}|\bm{x})$ for corresponding $\bm{x}$.
        
        First we note that the distribution $p(\bm{y}|\hat{\bm{x}})$ is given as
        \begin{equation} \label{self_consistent_equation_1}
            \begin{aligned}
                p(\bm{y}|\hat{\bm{x}}) &= \frac{1}{p(\hat{\bm{x}})}\sum_{\bm{x}}{p(\bm{y}|\bm{x})p(\hat{\bm{x}}|\bm{x})p(\bm{x})} \\
                &= \sum_{\bm{x}}{p(\bm{y}|\bm{x})p(\bm{x}|\hat{\bm{x}})}.
            \end{aligned}
        \end{equation}
        According to the Markov chain $\bm{Y} \leftrightarrow \bm{X} \leftrightarrow \hat{\bm{X}}$, we get $p(\hat{\bm{x}}|\bm{x},\bm{y}) = p(\hat{\bm{x}}|\bm{x})$ and $p(\bm{y}|\bm{x},\hat{\bm{x}}) = p(\bm{y}|\bm{x})$, then
        
        \begin{equation} \label{self_consistent_equation_2}
            p(\hat{\bm{x}}) = \sum_{\bm{x}}{p(\bm{x})p(\hat{\bm{x}}|\bm{x})},
        \end{equation}
        
        \begin{equation}
            \begin{aligned}
                p(\hat{\bm{x}}|\bm{y}) &=  \sum_{\bm{x}}{p(\hat{\bm{x}},\bm{x}|\bm{y})}  \\
                &=  \sum_{\bm{x}}{p(\hat{\bm{x}}|\bm{x},\bm{y})p(\bm{x}|\bm{y})}  \\
                &=  \sum_{\bm{x}}{p(\hat{\bm{x}}|\bm{x})p(\bm{x}|\bm{y})}.
            \end{aligned}
        \end{equation}
        From equations (\ref{self_consistent_equation_1}) and (\ref{self_consistent_equation_2}), we can get the derivatives w.r.t. $p(\hat{\bm{x}}|\bm{x})$
        
        \begin{eqnarray}
            \frac{\partial{p(\hat{\bm{x}})}}{\partial{p(\hat{\bm{x}}|\bm{x})}} &=& p(\bm{x}), \label{partial_1} \\
            \frac{\partial{p(\hat{\bm{x}}|\bm{y})}}{\partial{p(\hat{\bm{x}}|\bm{x})}} &=& p(\bm{x}|\bm{y}). \label{partial_2}
        \end{eqnarray}
        Taking derivatives of (\ref{eq_optimization_2}) with respect to $p(\hat{\bm{x}}|\bm{x})$ for given $\bm{x}$ and $\hat{\bm{x}}$
        
        \begin{equation} \label{derivative_1}
            \begin{aligned}
                \frac{\partial{\mathcal{L}(p(\hat{\bm{x}}|\bm{x})}}{\partial{p(\hat{\bm{x}}|\bm{x})}} 
                &= \lambda p(\bm{x}) \log \frac{p(\hat{\bm{x}}|\bm{x})}{p(\hat{\bm{x}})} + \lambda p(\bm{x})p(\hat{\bm{x}}|\bm{x})\frac{1}{p(\hat{\bm{x}}|\bm{x})} \\
                &\phantom{\;=} - \lambda \sum_{\bm{x}}{p(\bm{x})p(\hat{\bm{x}}|\bm{x})} \frac{p(\bm{x})}{p(\hat{\bm{x}})} + p(\bm{x})d_{RD}(\bm{x},\hat{\bm{x}}) \\
                &\phantom{\;=} - \beta \sum_{\bm{y}}{\frac{\partial{p(\hat{\bm{x}}|\bm{y})}}{\partial{p(\hat{\bm{x}}|\bm{x})}}p(\bm{y})[1+\log p(\hat{\bm{x}}|\bm{y})]} \\
                &\phantom{\;=} +  \beta \frac{\partial{p(\hat{\bm{x}}})}{\partial{p(\hat{\bm{x}}|\bm{x})}}[1+\log p(\hat{\bm{x}})] + r(\bm{x}). \\
            \end{aligned}
        \end{equation}
        Substitute (\ref{partial_1}) and (\ref{partial_2}) into (\ref{derivative_1})

        \begin{equation} \label{derivative_2}
            \begin{aligned}
                \frac{\partial{\mathcal{L}(p(\hat{\bm{x}}|\bm{x})}}{\partial{p(\hat{\bm{x}}|\bm{x})}} 
                &= \lambda p(\bm{x})\log\frac{p(\hat{\bm{x}}|\bm{x})}{p(\hat{\bm{x}})} + p(\bm{x})d_{RD}(\bm{x},\hat{\bm{x}})\\
                &\phantom{\;=} - \beta \sum_{\bm{y}}p(\bm{x},\bm{y})[1+\log p(\hat{\bm{x}}|\bm{y})] \\
                &\phantom{\;=} + \beta p(\bm{x})[1+\log p(\hat{\bm{x}})] + r(\bm{x}). \\
            \end{aligned}
        \end{equation}
        Rearranging this equation
        \begin{equation} \label{derivative_3}
            \begin{aligned}
                \frac{\partial{\mathcal{L}(p(\hat{\bm{x}}|\bm{x})}}{\partial{p(\hat{\bm{x}}|\bm{x})}} 
                &= \lambda p(\bm{x})\left[ \log\frac{p(\hat{\bm{x}}|\bm{x})}{p(\hat{\bm{x}})}+\lambda^{-1} d_{RD}(\bm{x},\hat{\bm{x}}) \right] \\
                &\phantom{\;=} - \beta p(\bm{x}) \left[ \sum_{\bm{y}}p(\bm{y}|\bm{x})\log {\frac{p(\bm{y}|\hat{\bm{x}})}{p(\bm{y})}}\right]  + r(\bm{x}).  \\
            \end{aligned}
        \end{equation}
        Notice that $\sum_{\bm{y}} p(\bm{y}|\bm{x}) \log \frac{p(\bm{y}|\bm{x})}{p(\bm{y})}=I(\bm{X};\bm{Y})$ is a function of $\bm{x}$ (independent of $\hat{\bm{x}}$), thus it can be absorbed into the multiplier $r(\bm{x})$. Introducing
        \begin{equation}
            \log \mu (\bm{x}) = \frac{1}{\lambda} \left( \frac{r(\bm{x})}{p(\bm{x})} - \beta \sum_{\bm{y}}p(\bm{y}|\bm{x}) \log \frac{p(\bm{y}|\bm{x})}{p(\bm{y})} \right).
        \end{equation}
        Finally, (\ref{derivative_3}) can be simplified to
        \begin{equation} \label{derivative_4}
            \begin{aligned}
                \frac{\partial{\mathcal{L}(p(\hat{\bm{x}}|\bm{x})}}{\partial{p(\hat{\bm{x}}|\bm{x})}} 
                &= \lambda p(\bm{x})\left[ \log\frac{p(\hat{\bm{x}}|\bm{x})}{p(\hat{\bm{x}})}+ \lambda^{-1}d_{RD}(\bm{x},\hat{\bm{x}}) \right] \\
                &\phantom{\;=} - \beta p(\bm{x}) \left[ \sum_{\bm{y}}p(\bm{y}|\bm{x})\log {\frac{p(\bm{y}|\hat{\bm{x}})}{p(\bm{y})}}\right] \\
                &\phantom{\;=} +\beta p(\bm{x}) \left[ \sum_{\bm{y}}{p({bm{y}|\bm{x}})\log \frac{p(\bm{y}|\bm{x})}{p(\bm{y})}} \right] \\
                &\phantom{\;=} - \beta p(\bm{x}) \left[ \sum_{\bm{y}}{p({bm{y}|\bm{x}})\log \frac{p(\bm{y}|\bm{x})}{p(\bm{y})}} \right] + r(\bm{x})  \\
                &= \lambda p(\bm{x})\left[ \log\frac{p(\hat{\bm{x}}|\bm{x})}{p(\hat{\bm{x}})}+ \lambda^{-1}d_{RD}(\bm{x},\hat{\bm{x}})\right] \\
                &\phantom{\;=} +\beta p(\bm{x}) \left[ \sum_{\bm{y}}{p({bm{y}|\bm{x}})\log \frac{p(\bm{y}|\bm{x})}{p(\bm{y}|\hat{\bm{x}})}}\right]\\ 
                &\phantom{\;=} + \lambda p(\bm{x}) \log \mu (\bm{x}), \\
            \end{aligned}
        \end{equation}
        We know that 
        \begin{equation}
            \begin{aligned}
                d_{T}(\bm{x},\hat{\bm{x}})
                &= D_{KL}[p(\bm{y}|\bm{x})||p(\bm{y}|\hat{\bm{x}}]\\
                &= \sum_{\bm{y}}{p({\bm{y}|\bm{x}})\log \frac{p(\bm{y}|\bm{x})}{p(\bm{y}|\hat{\bm{x}})}}.
            \end{aligned}
        \end{equation}
        Setting (\ref{derivative_4}) to zero,
        \begin{equation} \label{self_consistent_equation_3}
            \begin{aligned}
                p(\hat{\bm{x}}|\bm{x})=\frac{p(\hat{\bm{x}})e^{-\lambda ^{-1} d_{S}(\bm{x},\hat{\bm{x}})}}{\mu(\bm{x})},
            \end{aligned}
        \end{equation}
        where 
        \begin{equation}
            d_{S}(\bm{x},\hat{\bm{x}}) = d_{RD}(\bm{x},\hat{\bm{x}}) + \beta d_{T}(\bm{x},\hat{\bm{x}}).
        \end{equation}
        Since $\sum_{\hat{\bm{x}}}{p(\hat{\bm{x}}|\bm{x})=1}$, substitute it to (\ref{self_consistent_equation_3})
        \begin{equation}
            \mu(\bm{x})=\sum_{\hat{\bm{x}}}p(\hat{\bm{x}})e^{- \lambda^{-1} d_{S}(\bm{x},\hat{\bm{x}})}.
        \end{equation}
  \end{appendices}


\begin{thebibliography}{1}

\bibitem{shannon1948mathematical}
C.~E. Shannon, ``A mathematical theory of communication,'' \emph{The Bell
  system technical journal}, vol.~27, no.~3, pp. 379--423, 1948.

\bibitem{strinati20216g}
E.~C. Strinati and S.~Barbarossa, ``{6G} networks: Beyond shannon towards
  semantic and goal-oriented communications,'' \emph{Comm. Com. Inf. Sc.}, vol.
  190, p. 107930, 2021.

\bibitem{xie2021deep}
H.~Xie, Z.~Qin, G.~Y. Li, and B.-H. Juang, ``Deep learning enabled semantic
  communication systems,'' \emph{IEEE Trans. Signal Process.}, vol.~69, pp.
  2663--2675, 2021.

\bibitem{SPIC}
N.~Patwa, N.~Ahuja, S.~Somayazulu, O.~Tickoo, S.~Varadarajan, and S.~Koolagudi,
  ``Semantic-preserving image compression,'' in \emph{IEEE Int. Conf. Image
  Process. (ICIP)}, ABU Dhabi, United Arab Emirates, Oct. 2020, pp. 1281--1285.
  
\bibitem{Balle}
J.~Ball{\'e}, V.~Laparra, and E.~P. Simoncelli, ``End-to-end optimized image
  compression,'' in \emph{Proc. Int. Conf. Learn. Representations (ICLR)},
  Toulon, France, Apr. 2017.

\bibitem{cover1999elements}
T.~M. Cover, \emph{Elements of information theory}.\hskip 1em plus 0.5em minus
  0.4em\relax John Wiley \& Sons, 1999.

\bibitem{JPEG}
G.~K. Wallace, ``The {JPEG} still picture compression standard,'' \emph{IEEE
  transactions on consumer electronics}, vol.~38, no.~1, pp. xviii--xxxiv,
  1992.

\bibitem{JPEG2000}
M.~Rabbani and R.~Joshi, ``An overview of the {JPEG} 2000 still image
  compression standard,'' \emph{Signal processing: Image communication},
  vol.~17, no.~1, pp. 3--48, 2002.

\bibitem{BPG}
Bellard, Fabrice, ``{BPG} Image Format,'' 2021, [Online].
  Avaible: https://bellard.org/bpg/.

\bibitem{HEVC}
G.~J. Sullivan, J.-R. Ohm, W.-J. Han, and T.~Wiegand, ``Overview of the high
  efficiency video coding ({HEVC}) standard,'' \emph{IEEE Trans. Circuits Syst.
  Video Technol.}, vol.~22, no.~12, pp. 1649--1668, 2012.

\bibitem{VVC}
B.~Benjamin, C. Jianle, L. Shan and W. Ye-Kui, ``{JVET-S2001} Versatile Video Coding (Draft 
  10),'' \emph{Proc. Joint Video Exploration Team (JVET) ofITU-T SG 16 WP 3 and ISO/IEC JTC 1/SC 29/WG 11}, 2014.

\bibitem{Theis}
L.~Theis, W.~Shi, A.~Cunningham, and F.~Husz{\'a}r, ``Lossy image compression
  with compressive autoencoders,'' in \emph{Proc. Int. Conf. Learn.
  Representations (ICLR)}, Toulon, France, Apr. 2017.

\bibitem{Toderici}
G.~Toderici, S.~M. O'Malley, S.~J. Hwang, D.~Vincent, D.~Minnen, S.~Baluja,
  M.~Covell, and R.~Sukthankar, ``Variable rate image compression with
  recurrent neural networks,'' in \emph{Proc. Int. Conf. Learn. Representations
  (ICLR)}, SAN Juan, Puerto Rico, May. 2016.

\bibitem{ScaleHyperprior}
J.~Ball{\'e}, D.~Minnen, S.~Singh, S.~J. Hwang, and N.~Johnston, ``Variational
  image compression with a scale hyperprior,'' in \emph{Proc. Int. Conf. Learn.
  Representations (ICLR)}, Vancouver, Canada, Apr. 2018.


\bibitem{CAECheng}
Z.~Cheng, H.~Sun, M.~Takeuchi, and J.~Katto, ``Learned image compression with
  discretized gaussian mixture likelihoods and attention modules,'' in
  \emph{Proc. IEEE/CVF Conf. Comput. Vis. Pattern Recognit (CVPR)}, Seattle,
  United States, Jun. 2020, pp. 7939--7948.

\bibitem{DIC}
Z.~Yang, Y.~Wang, C.~Xu, P.~Du, C.~Xu, C.~Xu, and Q.~Tian, ``Discernible image
  compression,'' in \emph{Proc. 28th ACM Int. Conf. Multimedia (ACMMM)},
  Westminster, United States, Oct. 2020, pp. 1561--1569.

\bibitem{IB1999}
N.~Tishby, F.~C. Pereira, and W.~Bialek, ``The information bottleneck method,''
  in \emph{Proc. 37th Annual Allerton Conf. Commun. Control Comput.}, Allerton
  House, Monticello, Illinois, Sep. 1999.

\bibitem{SCAITasks}
Y.~Yang, C.~Guo, F.~Liu, C.~Liu, L.~Sun, Q.~Sun, and J.~Chen, ``Semantic
  communications with ai tasks,'' 2021, \textit{arXiv:2109.14170}. [Online].
  Avaible: https://arxiv.org/abs/2109.14170.

\bibitem{DeepSIC}
S.~Luo, Y.~Yang, Y.~Yin, C.~Shen, Y.~Zhao, and M.~Song, ``{DeepSIC}: Deep
  semantic image compression,'' in \emph{Proc. Int. Conf. on Neural Inf.
  Process. (ICONIP)}, Siem Reap, Cambodia, Dec. 2018, pp. 96--106.
  
\bibitem{qin2021survey}
Z.~Qin, X.~Tao, J.~Lu, and G.~Y. Li, ``Semantic communications: Principles and
  challenges,'' 2021, \textit{arXiv:2201.01389}. [Online]. Avaible:
  https://arxiv.org/abs/2201.01389.
  
\bibitem{CLUB}
P.~Cheng, W.~Hao, S.~Dai, J.~Liu, Z.~Gan, and L.~Carin, ``{CLUB}: A contrastive
  log-ratio upper bound of mutual information,'' in \emph{Proc. Int. Conf.
  Mach. Learning (ICML)}, Vienna, Austria, Jul. 2020, pp. 1779--1788.

\bibitem{Carnap}
R.~Carnap, Y.~Bar-Hillel \emph{et~al.}, ``An outline of a theory of semantic
  information,'' 1952.

\bibitem{Situation}
J.~Barwise and J.~Perry, ``Situations and attitudes,'' \emph{J. Philos.},
  vol.~78, no.~11, pp. 668--691, 1981.

\bibitem{Strong}
L.~Floridi, ``Outline of a theory of strongly semantic information,''
  \emph{Minds and machines}, vol.~14, no.~2, pp. 197--221, 2004.

\bibitem{Zadeh_1}
J.~Goguen, La zadeh. ``fuzzy sets'', information and control, vol. 8 (1965), pp.
  338--353.-la zadeh. similarity relations and fuzzy orderings. information
  sciences, vol. 3 (1971), pp. 177--200. \emph{J. Symb. Log.}, vol.~38,
  no.~4, pp. 656--657, 1973.

\bibitem{Zadeh_2}
L.~A. Zadeh, ``Probability measures of fuzzy events,'' \emph{J. of Math. Anal.
  Appl}, vol.~23, no.~2, pp. 421--427, 1968.
  
\bibitem{6GSemantic}
P.~Zhang, W.~Xu, H.~Gao, K.~Niu, X.~Xu, X.~Qin, C.~Yuan, Z.~Qin, H.~Zhao,
  J.~Wei \emph{et~al.}, ``Toward wisdom-evolutionary and primitive-concise
  {6G}: A new paradigm of semantic communication networks,'' \emph{Eng.},
  vol.~8, pp. 60--73, 2022.


\bibitem{Speech_Recognition}
Z.~Weng, Z.~Qin, and G.~Y. Li, ``Semantic communications for speech
  recognition,'' 2021, \textit{arXiv:2107.11190}. [Online]. Avaible:
  https://arxiv.org/abs/2107.11190.

\bibitem{IB2015}
N.~Tishby and N.~Zaslavsky, ``Deep learning and the information bottleneck
  principle,'' in \emph{Proc. IEEE Inf. Theory Workshop (ITW)}, Israel, Apr.
  2015, pp. 1--5.

\bibitem{MINE}
M.~I. Belghazi, A.~Baratin, S.~Rajeshwar, S.~Ozair, Y.~Bengio, A.~Courville,
  and D.~Hjelm, ``Mutual information neural estimation,'' in \emph{Proc. Int.
  Conf. Mach. Learning (ICML)}, Stockholm, Sweden, Jun. 2018, pp. 531--540.

\bibitem{JSCC}
E.~Bourtsoulatze, D.~B. Kurka, and D.~G{\"u}nd{\"u}z, ``Deep joint
  source-channel coding for wireless image transmission,'' \emph{IEEE Trans. on
  Cogn. Commun. Netw.}, vol.~5, no.~3, pp. 567--579, 2019.

\bibitem{JSCC-f}
D.~B. Kurka and D.~G{\"u}nd{\"u}z, ``DeepJSCC-f: Deep joint source-channel
  coding of images with feedback,'' \emph{IEEE J. Sel. Areas Inf. Theory},
  vol.~1, no.~1, pp. 178--193, 2020.

\bibitem{JointTR}
C.-H. Lee, J.-W. Lin, P.-H. Chen, and Y.-C. Chang, ``Deep learning-constructed
  joint transmission-recognition for internet of things,'' \emph{IEEE Access},
  vol.~7, pp. 76\,547--76\,561, 2019.

\bibitem{JSCCRetrieval}
M.~Jankowski, D.~G{\"u}nd{\"u}z, and K.~Mikolajczyk, ``Deep joint
  source-channel coding for wireless image retrieval,'' in \emph{Proc.
  ICASSP-IEEE Int. Conf. Acoust., Speech Signal Process. (ICASSP)}, Barcelona,
  Spain, May. 2020, pp. 5070--5074.

\bibitem{ImageRetrieval}
M.~Jankowski, D.~G{\"u}nd{\"u}z, and K.~Mikolajczyk, ``Wireless image retrieval
  at the edge,'' \emph{IEEE J. Sel. Areas Commun.}, vol.~39, no.~1, pp.
  89--100, 2020.

\bibitem{multi_modal_1}
H.~Xie, Z.~Qin, and G.~Y. Li, ``Task-oriented semantic communications for
  multimodal data,'' 2021, \textit{arXiv:2108.07357}. [Online]. Avaible:
  https://arxiv.org/abs/2108.07357.

\bibitem{multi_modal_2}
H.~Xie, Z.~Qin, X.~Tao, and K.~B. Letaief, ``Task-oriented multi-user semantic
  communications,'' 2021, \textit{arXiv:2112.10255}. [Online]. Avaible:
  https://arxiv.org/abs/2112.10255.

\bibitem{Semantic_speech}
Z.~Weng and Z.~Qin, ``Semantic communication systems for speech transmission,''
  \emph{IEEE J. Sel. Areas Commun.}, vol.~39, no.~8, pp. 2434--2444, 2021.

\bibitem{he2021masked}
K.~He, X.~Chen, S.~Xie, Y.~Li, P.~Doll{\'a}r, and R.~Girshick, ``Masked
  autoencoders are scalable vision learners,'' 2021, \textit{arXiv:2111.06377}.
  [Online]. Avaible: https://arxiv.org/abs/2111.06377.



\bibitem{ImportantMap}
M.~Li, W.~Zuo, S.~Gu, D.~Zhao, and D.~Zhang, ``Learning convolutional networks
  for content-weighted image compression,'' in \emph{Proc. IEEE Conf. Comput.
  Vis. Pattern Recognit. (CVPR)}, Salt Lake City, USA, Jun. 2018, pp.
  3214--3223.

\bibitem{ROI}
C.~Cai, L.~Chen, X.~Zhang, and Z.~Gao, ``End-to-end optimized {ROI} image
  compression,'' \emph{IEEE Trans. Image Process.}, vol.~29, pp. 3442--3457,
  2019.
  
\bibitem{DeepVision}
J.~Chai, H.~Zeng, A.~Li, and E.~W. Ngai, ``Deep learning in computer vision: A
  critical review of emerging techniques and application scenarios,''
  \emph{Mach. Learn. Appl.}, vol.~6, p. 100134, 2021.

\bibitem{App1}
N.~Tishby and N.~Slonim, ``Data clustering by markovian relaxation and the
  information bottleneck method,'' in \emph{Proc. Adv. Neural Inf. Process.
  Syst. (NIPS)}, Denver, CO, USA, Nov. 2000.

\bibitem{App2}
N.~Friedman, O.~Mosenzon, N.~Slonim, and N.~Tishby, ``Multivariate information
  bottleneck,'' 2013, \textit{arXiv:1301.2270}. [Online]. Avaible:
  https://arxiv.org/abs/1301.2270.

\bibitem{App3}
N.~Slonim, G.~S. Atwal, G.~Tka{\v{c}}ik, and W.~Bialek, ``Information-based
  clustering,'' \emph{Proc. Natl. Acad. Sci.}, vol. 102, no.~51, pp.
  18\,297--18\,302, 2005.

\bibitem{IB2017}
R.~Shwartz-Ziv and N.~Tishby, ``Opening the black box of deep neural networks
  via information,'' 2017, \textit{arXiv:1703.00810}. [Online]. Available:
  https://arxiv.org/abs/1703.00810.

\bibitem{VIB}
A.~A. Alemi, I.~Fischer, J.~V. Dillon, and K.~Murphy, ``Deep variational
  information bottleneck,'' Apr. 2017.

\bibitem{RFBNet}
S.~Liu, D.~Huang \emph{et~al.}, ``Receptive field block net for accurate and
  fast object detection,'' in \emph{Proc. Eur. Conf. Comput. Vision (ECCV)},
  Munich, Germany, Sep. 2018, pp. 385--400.

\bibitem{GDN}
J.~Ball{\'e}, V.~Laparra and E. Simoncelli, ``Density modeling of images using a generalized normalization transformation,'' 2015, \textit{arXiv:1511.06281}. [Online]. Available:
  https://arxiv.org/abs/1511.06281.

\bibitem{STL_10}
A.~Coates, Ng.~A and H. Lee \emph{et~al.}, ``An analysis of single-layer networks in unsupervised feature learning,'' in \emph{Proc. 14th Int. Conf. Artif. Intell. Stat. (AISTATS)},
  Ft. Lauderdale, FL, USA, 2011.

\bibitem{Pascal_VOC}
M.~Everingham, L.~Van Gool. Atwal, C.~Williams, et.al, ``The pascal visual object classes (voc) challenge,'' \emph{Int. J. Comput. Vis.}, vol. 88, no.~2, pp.
  303--338, 2010.

\end{thebibliography}
\end{document}